\documentclass[10pt,twocolumn,letterpaper]{article}
\usepackage{preprint}

\usepackage{cuted} 

\usepackage{booktabs}
\usepackage{pifont}
\definecolor{encodinggreen}{HTML}{228B22}
\definecolor{encodingred}{HTML}{D32F2F}
\newcommand{\encodingcheck}{\textcolor{encodinggreen}{\ding{51}}}
\newcommand{\encodingcross}{\textcolor{encodingred}{\ding{55}}}

\usepackage{graphicx}       
\usepackage{colortbl,xcolor}
\definecolor{firstbest}{HTML}{C6EFCE}
\definecolor{secondbest}{HTML}{f3f19b}
\definecolor{thirdbest}{HTML}{f7ce7c}
\usepackage{currfile} 

\usepackage{caption} 

\definecolor{linkblue}{rgb}{0.21,0.49,0.74}
\usepackage[breaklinks,colorlinks,allcolors=linkblue]{hyperref}
\hypersetup{
  pdftitle={TetraCoder: Adaptive Tetrahedral Encoding for Neural Fields},
  pdfauthor={Daniele Della Pietra, Antonella Rech, Nicola Conci, Nicola Garau},
  pdfsubject={Neural field representations},
  pdfkeywords={TetraCoder, neural fields, tetrahedral encoding}
}

\usepackage{multirow}
\newcommand{\ourmethod}{\textsc{TetraCoder}}
\newcommand{\hash}{\textsc{Hash}}
\newcommand{\thash}{\textsc{THash}}

\newcommand{\mesh}{\textsc{Mesh}}

\newcommand{\taskone}{Toy 2D image fitting}
\newcommand{\tasktwo}{3D point cloud to mesh}
\newcommand{\taskthree}{3D scene reconstruction from images}
\newcommand{\taskfour}{3D scene editing}

\title{\ourmethod: Adaptive Tetrahedral Encoding for Neural Fields}
\author{Daniele Della Pietra \quad Antonella Rech \quad Nicola Conci \quad Nicola Garau\\
University of Trento, Trento, Italy
}
\date{}

\begin{document}
\maketitle
\begin{strip}
    \centering
    \includegraphics[width=.33\linewidth]{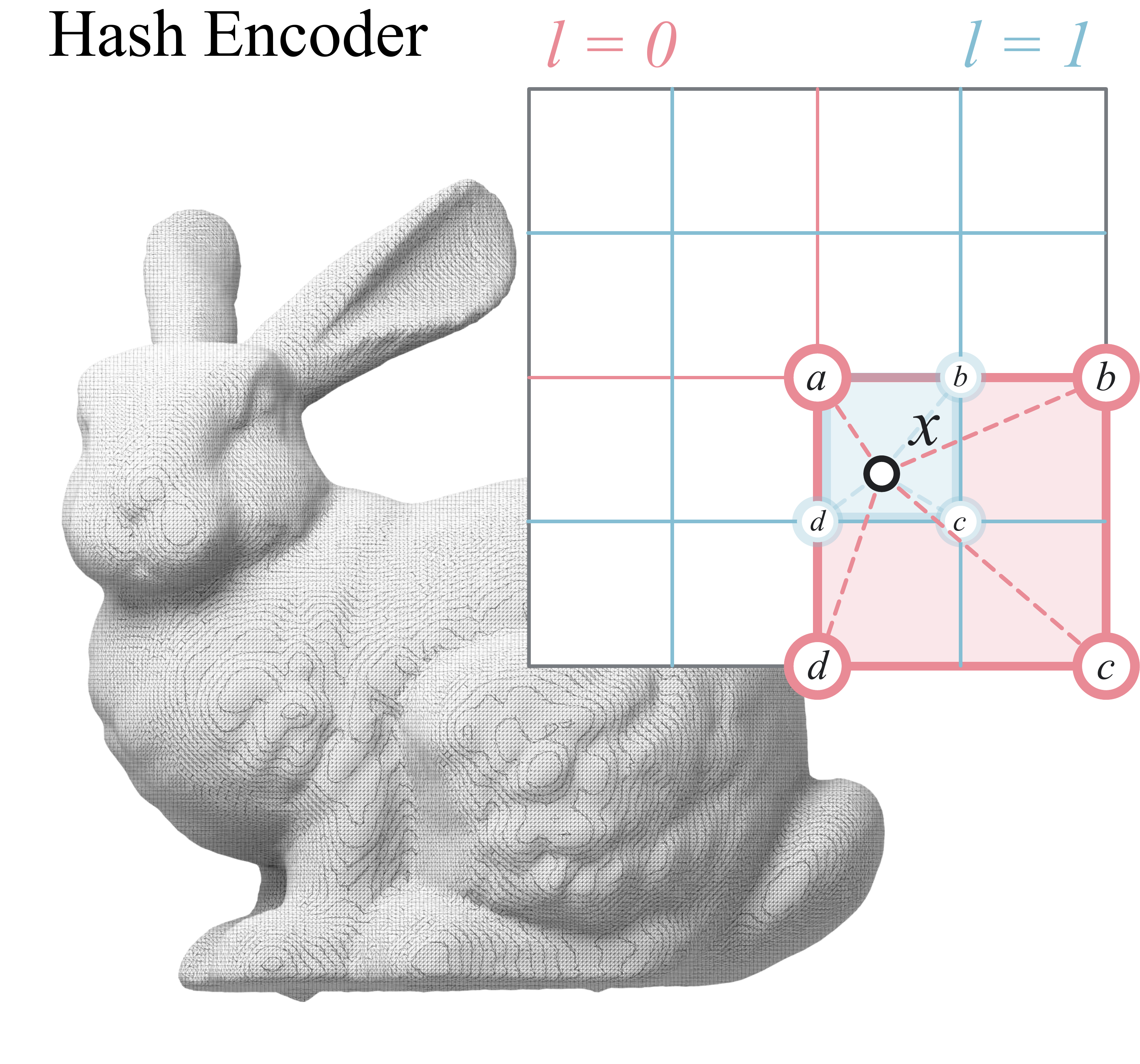}\hfill
    \includegraphics[width=.33\linewidth]{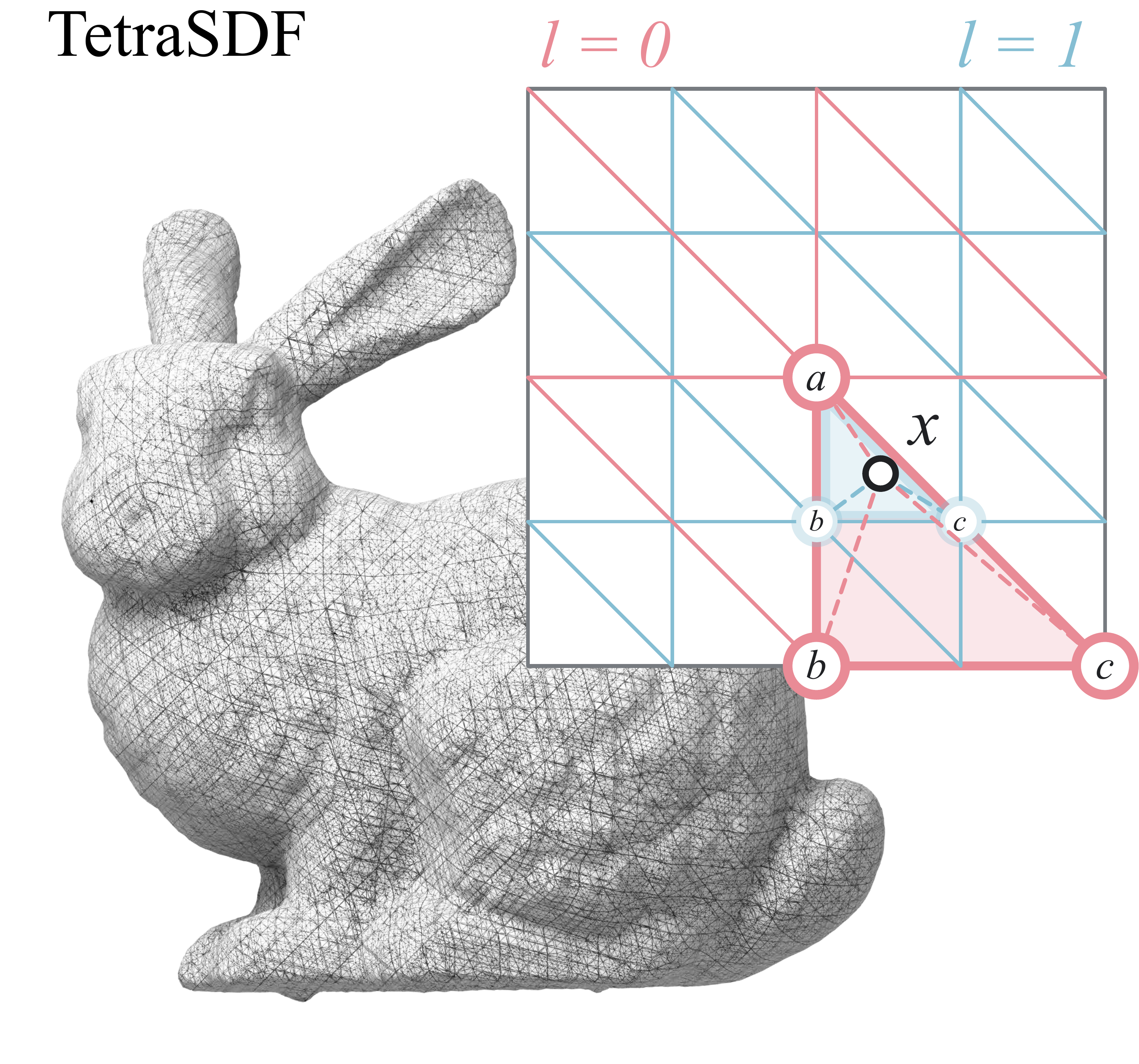}\hfill
    \includegraphics[width=.33\linewidth]{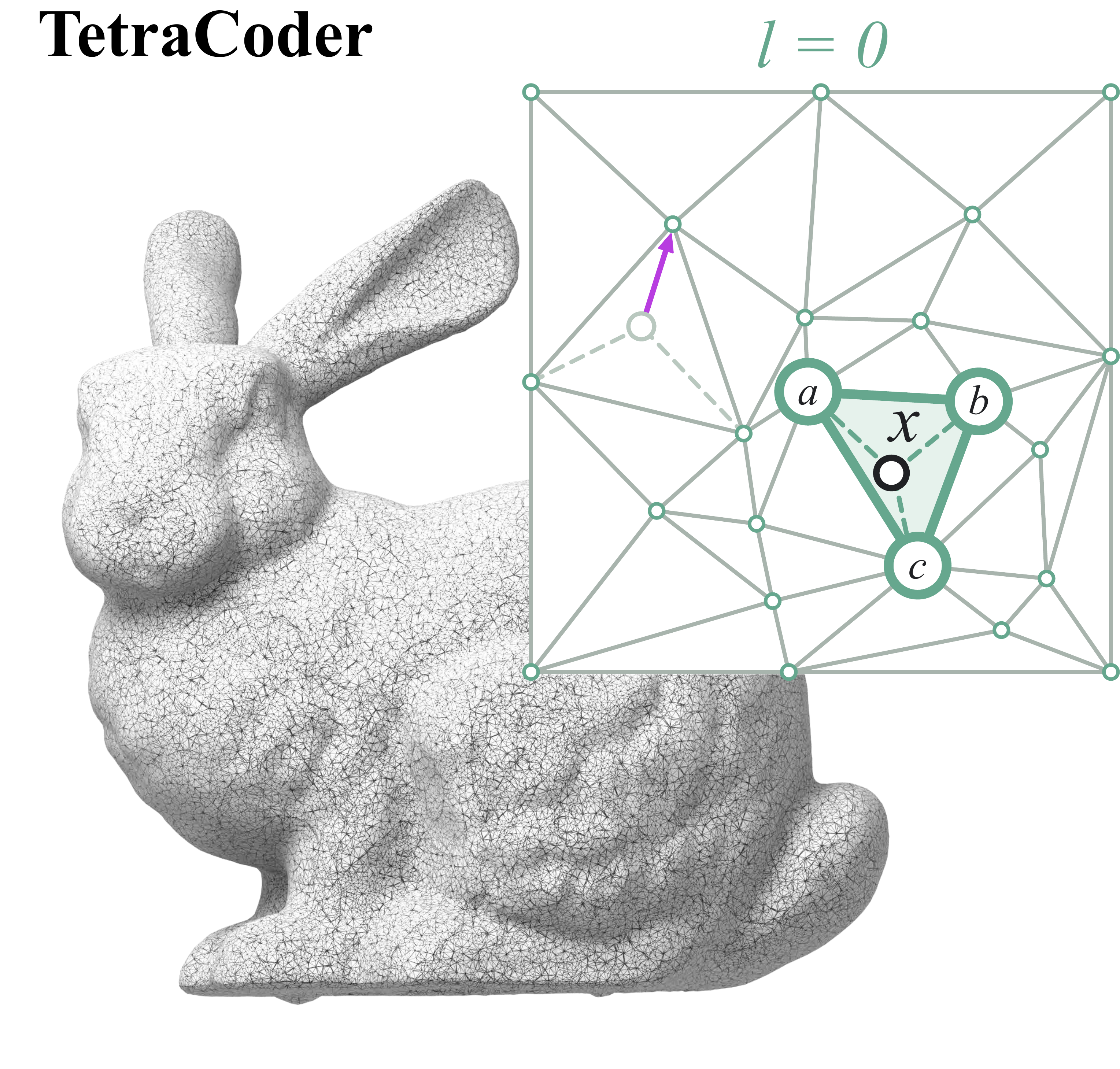}
    \captionsetup{hypcap=false}
    \captionof{figure}{We present \textbf{\ourmethod}, an adaptive tetrahedral encoding for neural fields. Multiresolution hash encoding (\textbf{left}) interpolates features on fixed spatial grids at multiple resolutions. Its trilinear interpolation does not produce a piecewise-affine encoding, making analytic mesh extraction impractical. To address that, tetrahedral multiresolution hash encoding (\textbf{center}) subdivides these grids into fixed tetrahedra and uses barycentric interpolation to obtain a piecewise-affine encoding. \ourmethod{} (\textbf{right}) instead uses a single-level, hash-free encoding defined on an adaptive tetrahedral volumetric mesh. Combined with a piecewise-affine decoder, this representation enables fast analytic mesh retrieval. Unlike other encodings, \ourmethod{} supports novel view synthesis, scene reconstruction, 3D appearance and geometry editing within a unified framework. Visit our project page at \url{https://mmlab-cv.github.io/TetraCoder/}.}
    \captionsetup{hypcap=true}
    \label{fig:\currfilebase}%
\end{strip}

\begin{abstract}

Implicit and explicit representations of geometry are central in computer graphics applications and come with pros and cons. With surface sampling being faster on explicit representations and spatial queries being faster on implicit representations, finding efficient conversions between them is a highly desirable property.
We present \ourmethod{}, a single-resolution hash-free adaptive tetrahedral encoding that links an explicit tetrahedral spatial encoder with an underlying implicit signed distance field (SDF). The resulting hybrid implicit-explicit space is continuous piecewise affine (CPWA) by design, yielding an exact correspondence between the implicit SDF and the explicit analytic mesh.
\ourmethod{}'s design comes with additional perks, in particular: (a) it scales better than multiresolution encoding methods due to its single-resolution optimizable grid, (b) unlike hash-based methods, it is not affected by collisions and (c) it enables fast tetrahedral ray-tracing, geometry and appearance editing, directly through its adaptive tetrahedral scaffold.
We demonstrate the versatility of \ourmethod{} by showing an improvement over state-of-the-art methods on multiple tasks: 2D/3D fitting from images and point clouds, 3D novel view synthesis, 3D reconstruction, and geometry/appearance editing.

\end{abstract}
\section{Introduction}
\label{sec:intro}

A neural field is a function $f_\theta$ represented by a neural network that maps coordinates to quantities such as color, density, or signed distance. Formally,
\begin{equation}
    \mathbf{y} = f_{\theta}(\mathbf{x}),
\end{equation}
where $\mathbf{x}$ denotes the input coordinates, $\mathbf{y}$ the predicted field values, and $\theta$ the learned parameters.

Neural fields support learning continuous geometry and view-dependent appearance from observations~\cite{mildenhall2020nerf,Neus,wang2023neus2}.
Using these reconstructions for geometric processing, simulation, and editing also benefits from explicit surfaces.
For an SDF, obtaining such a surface requires extracting its zero-level set.
Sampling-based mesh retrieval approximates this surface, with accuracy depending on the sampling resolution~\cite{lorensen1987marching}.
Continuous piecewise-affine (CPWA) fields enable analytic retrieval of the learned surface~\cite{lei2020analytic,tetrasdf}, establishing a direct correspondence between implicit and explicit geometry with no sampling errors.
The spatial encoding therefore influences not only fitting accuracy but also how the learned surface can be extracted and edited.

Existing encodings come with different compromises.
Multiresolution hash grids capture details efficiently~\cite{muller2022instant}, but hash collisions can couple distant regions, and trilinear interpolation does not preserve piecewise-affine structure.
Tetrahedral encodings enable analytic mesh retrieval provided that the downstream decoders don't introduce nonlinearities, yet overlapping grid levels require costly geometric intersections~\cite{tetrasdf}.
Fixed tetrahedral encodings~\cite{kulhanek2023tetra} cannot adapt their spatial domains during training, while mesh-based encodings~\cite{yang2022neumesh,bao2025neumeshplusplus} depend on reconstructed surfaces.

We introduce \ourmethod{}, a single-level, hash-free adaptive tetrahedral encoding (Fig.~\ref{fig:teaser}) that establishes a direct correspondence between the learned implicit surface and its explicit representation, while enabling editing through the encoder grid itself.
Barycentric feature interpolation and ReLU geometry decoding preserve the CPWA structure, while vertex insertion, pruning, and movement adjust the density and placement of grid vertices to capture fine geometric and appearance details.
The single grid enables analytic mesh retrieval without cross-level intersections, supports tetrahedral ray traversal, and provides geometry and appearance editing through its vertices.

We evaluate \ourmethod{} on image fitting, mesh reconstruction from point clouds, joint novel-view synthesis and mesh reconstruction from RGB images, and scene editing.
Our results demonstrate better fitting with fewer parameters for the first two tasks and the best average geometry and rendering scores among evaluated baselines on NeRF Synthetic and DTU.
For mesh reconstruction from geometric data, \ourmethod{} reduces mean analytic extraction time by a factor of approximately $6.4$ compared with TetraSDF~\cite{tetrasdf}. \textbf{Our contributions are:}
\begin{itemize}
    \item We present a new general encoder for neural fields that matches or outperforms multiresolution hash encoding on a series of tasks, with less trainable parameters.
    \item Our encoder stores radiance and surface features locally in a single-level foam of adaptive tetrahedra, addressing the issues of fixed spatial grid and hash collisions.
    \item \ourmethod{} builds a direct correspondence between the learned implicit surface and its explicit surface representations, enabling analytic mesh retrieval.
\end{itemize}


\section{Related Work}
\label{sec:related_work}


\paragraph{Spatial encodings for neural fields.}
Neural networks often learn low-frequency components before capturing fine details, a tendency known as spectral bias~\cite{rahaman2019spectral}.
For neural fields, this motivates enriching the decoder's input with Fourier~\cite{tancik2020fourier} or local learned features to better capture spatial detail.
Local capacity can be concentrated where it is needed~\cite{govindarajan2024lagrangian} through sparse voxel refinement and pruning~\cite{liu2020neural}, hierarchical octrees~\cite{takikawa2021nglod}, or block decompositions that adapt to signal complexity~\cite{martel2021acorn}.
Multiresolution grids offer a complementary way to capture detail, with hashing reducing their storage cost~\cite{muller2022instant}.
Beyond storage, the interpolation rule determines how features are interpolated in space: barycentric interpolation on permutohedral lattices~\cite{rosu2023permutosdf} or tetrahedral hash grids~\cite{tetrasdf} preserve a continuous piecewise-affine feature field.
Features can also follow reconstructed geometry through point-supported fields~\cite{xu2022pointnerf} or interpolation over a tetrahedralization derived from a point cloud~\cite{kulhanek2023tetra}.
These choices present different constraints: hashing can couple features at unrelated locations~\cite{muller2022instant}, fixed tetrahedralizations cannot redistribute vertices during training~\cite{kulhanek2023tetra}, and multiresolution tetrahedral encodings introduce additional extraction costs by requiring intersections between tetrahedra across multiple grid levels~\cite{tetrasdf}.
\ourmethod{} addresses these constraints by storing features without hashing on a single tetrahedral grid that adapts through vertex movement, insertion, and pruning.


\paragraph{Neural reconstruction and novel-view synthesis.}
Implicit neural representations learn continuous fields, allowing novel views to be synthesized by volume rendering using density and view-dependent radiance~\cite{mildenhall2020nerf}.
Since density alone does not uniquely define a surface, geometry can instead be represented through occupancy~\cite{oechsle2021unisurf} or an SDF whose zero-level set defines the surface~\cite{yariv2021volume,Neus}.
Combining SDF-based reconstruction with multiresolution hash encoding improves training efficiency and captures finer geometric detail~\cite{wang2023neus2,li2023neuralangelo}.

Scene properties can also be stored directly on explicit primitives, such as voxels~\cite{fridovich2022plenoxels}, 3D Gaussians~\cite{kerbl2023gaussians} and surfels~\cite{10.1145/344779.344936}.
Using planar Gaussians~\cite{Huang2DGS2024} or aligning them with a surface~\cite{guedon2024sugar} makes these primitives better suited to surface reconstruction.
Other methods optimize a mesh in-the-loop together with 3D Gaussians~\cite{guedon2025milo} or by constructing a connected mesh from triangle splats and progressively making its faces opaque~\cite{held2026meshsplatting}.
Foam-based methods partitions space into Voronoi regions~\cite{RadiantFoam, govindarajan2026powerfoam, taveira2026differentiablevoronoiraytracing} or tetrahedra~\cite{mai2026radiance}, simplifying ray traversal.
\ourmethod{} combines the connected spatial structure of foam-based rendering with SDF-based volume rendering by decoding features interpolated over an adaptive tetrahedral grid.
Neural decoders map the vertex features interpolated by \ourmethod{} to continuous SDF and radiance fields, allowing geometry and appearance to vary within each tetrahedron.


\paragraph{Surface representations and analytic extraction.}
Analytic surface mesh retrieval instead aims to retrieve the mesh that exactly matches the SDF.
Marching Cubes~\cite{lorensen1987marching} extracts an isosurface from sampled grid values, while Dual Contouring~\cite{ju2002dual} uses edge intersections and surface normals to guide vertex placement and capture sharp features.
In tetrahedral grids, linear interpolation between scalar vertex values locates edge crossings that are connected into triangles by Marching Tetrahedra~\cite{doi1991tetrahedral}.
Making this extraction differentiable allows both vertex positions and SDF values to be optimized through losses on the resulting mesh~\cite{shen2021dmtet}.
Adaptive Delaunay grids extend this flexibility through error-driven resampling~\cite{binninger2025tetweave} and can also be paired with a separate hash-encoded SDF for image-based reconstruction~\cite{weidenfeller2026adele}.
Combining tetrahedral SDFs with cell rasterization further integrates differentiable extraction into rendering-based optimization~\cite{cui2026sdfraster}.
For a linearly interpolated SDF,~\cite{doi1991tetrahedral} recovers the represented surface exactly; a neural field, however, can contain details inside each tetrahedron that interpolation between its four vertex samples does not capture.

For ReLU MLPs, analytic extraction addresses this issue by identifying the regions in which the network is affine~\cite{lei2020analytic,berzins2023polyhedral,stippel2025marching}.
When a ReLU decoder operates on spatially interpolated features, the interpolation must also preserve piecewise-affine structure for  analytic extraction to apply.
Trilinear interpolation generally breaks this structure, motivating Eikonal-based planarity analysis and approximate intersection calculations~\cite{kim2024tropical}, whereas barycentric interpolation preserves it when combined with a ReLU decoder~\cite{tetrasdf}.
However, using multiple tetrahedral grid resolutions requires resolving intersections between the grids and the decoder's activation regions, making analytic extraction increasingly costly as the number of levels grows~\cite{tetrasdf}.
\ourmethod{} follows this analytic approach on a single adaptive tetrahedral grid, resolving the decoder's ReLU activation regions to exactly recover the learned zero-level set and allowing it to scale better compared to current state-of-the-art methods~\cite{tetrasdf,stippel2025marching}.


\paragraph{Editable neural fields.}
Editing a learned field often relies on an explicit geometric control structure, such as a mesh, rig, or cage.
Mesh deformations can be transferred to a pretrained radiance field through a tetrahedral proxy~\cite{yuan2022nerfediting}, with enclosing cages extending this idea to coarse volumetric controls~\cite{xu2022deforming} and interactive manipulation~\cite{jambon2023nerfshop}.
An alternative is to make the control surface part of the representation itself by storing separate geometry and appearance codes at mesh vertices, supporting both deformation and local texture editing~\cite{yang2022neumesh}.
Local parameterizations and semantic guidance extend these controls and improve interaction efficiency~\cite{bao2025neumeshplusplus}.
Editing through a proxy relies on a separate deformation mapping, whereas storing codes on a surface mesh ties both the representation and its controls to that surface.
\ourmethod{} uses the tetrahedral encoder itself, without needing an additional structure, as the editing domain: cage deformations move its vertices directly, while appearance edits modify its local features with the trained decoders fixed.

\begin{figure}
    \centering
    \includegraphics[width=\linewidth]{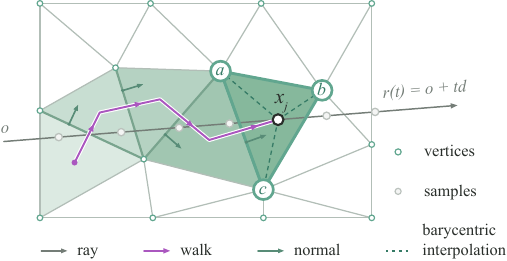}
    \caption{\textbf{Tetrahedral ray tracing.} For a sample $x_j$ along the ray $r(t)=o+td$, an adjacency walk (purple) locates the containing cell. Shading highlights the visited cells, while green arrows indicate face normals. Dashed lines illustrate barycentric interpolation of vertex features within the target triangle $(a,b,c)$. Gray markers denote additional samples along the ray.}
    \label{fig:tetrahedral_traversal}
\end{figure}

\section{Method}
\label{sec:method}
This section is split into two parts. Section \ref{sec:adaptive_encoder} describes the core structure, supported operations, and key properties of \ourmethod{}. All the presented details are meant to be agnostic to the task and can apply to other tasks that go beyond the scope of this paper. Section \ref{sec:tasks} presents four scenarios that rely on our encoder to tackle increasingly difficult tasks. 
Task \textbf{T1} (\taskone) shows that \ourmethod{} has enough capacity to fit an image into the weights of a small MLP, starting from 2D coordinates alone. 
Task \textbf{T2} (\tasktwo) extends this idea to the 3D domain, where the MLP now stores a SDF in its weights. Being \ourmethod{} CPWA by design, we showcase its fast analytic mesh retrieval capabilities.
Task \textbf{T3} (\taskthree) demonstrates how \ourmethod{} jointly learns geometry and view-dependent appearance from a collection of RGB images.
Its single-level adaptive tetrahedral grid supports both novel-view synthesis and explicit mesh reconstruction within the same representation, without additional spatial discretization, auxiliary 3D scaffolds, or mesh-in-the-loop learning.
Task \textbf{T4} (\taskfour) proves this idea even further, showcasing the link provided by \ourmethod{} between the implicit SDF and explicit analytic mesh. Displacing the tetrahedral foam directly deforms geometry, while its hash-free design enables localized appearance edits without hash collisions causing changes in distant regions, thus allowing \ourmethod{} to become the editing domain itself. More information about pruning, densification, and regularization hyperparameters is available in the supplementary material.

\subsection{Adaptive tetrahedral encoder}
\label{sec:adaptive_encoder}
\paragraph{Grid and interpolation.}
\ourmethod{} partitions a three-dimensional spatial domain using a single tetrahedral grid $\mathcal{T}$, where the single tetrahedron is defined as $\tau\in\mathcal{T}$.
The vertices of this grid are the corner points of its tetrahedra.
Each grid vertex $v$ stores a position $\mathbf{p}_v\in\mathbb{R}^{3}$ and a learned feature vector $\mathbf{f}_v$.
To encode an input point $\mathbf{x}\in\mathbb{R}^{3}$, we first locate the tetrahedron $\tau$ containing it.
Let $\operatorname{vert}(\tau)$ denote its four grid vertices.
For each vertex $v$, we form $\tau_{v\leftarrow\mathbf{x}}$ by replacing its position $\mathbf{p}_v$ with $\mathbf{x}$ while keeping the other three vertices fixed.
The barycentric weight of $v$ is the volume of this tetrahedron relative to the original volume:
\begin{equation}
    \lambda_v
    =\frac{V(\tau_{v\leftarrow\mathbf{x}})}
           {V(\tau)},
    \qquad v\in\operatorname{vert}(\tau),
    \label{eq:encoder_barycentric}
\end{equation}
where $V$ denotes tetrahedral volume.
For a point inside $\tau$, the four resulting volumes sum to the original volume, so the weights are nonnegative and sum to one.
We then use these weights to interpolate the learned vertex features, obtaining the encoded feature vector:
\begin{equation}
    \mathbf{e}(\mathbf{x})
    =\sum_{v\in\operatorname{vert}(\tau)}
    \lambda_v\,\mathbf{f}_v.
    \label{eq:encoder_interpolation}
\end{equation}
\paragraph{CPWA structure and locality.}
The construction above gives \ourmethod{} two fundamental properties: a CPWA encoding and hash-free feature storage.

Continuity follows from interpolation over shared grid vertices.
On a face between adjacent tetrahedra, the weights of the opposite vertices vanish, while the remaining weights coincide with the barycentric coordinates of that face.
Both tetrahedra therefore interpolate the same features with the same weights, producing identical values at their interface.

Piecewise affinity follows from the volume-based definition of barycentric weights.
Within a fixed tetrahedron $\tau$, each volume in the numerator of Eq.~\eqref{eq:encoder_barycentric} has a fixed base: the face opposite vertex $v$.
For $\mathbf{x}$ inside $\tau$, the height above this face varies affinely with $\mathbf{x}$.
Since the denominator is constant, each barycentric weight is affine in $\mathbf{x}$.
The encoding is a weighted sum of fixed vertex feature vectors, so it can be expressed by one affine function per tetrahedron:
\begin{equation}
    \mathbf{e}(\mathbf{x})
    =\mathbf{A}_{\tau}\mathbf{x}+\mathbf{b}_{\tau},
    \qquad \mathbf{x}\in\tau,
    \label{eq:encoder_affine}
\end{equation}
Here, the matrix $\mathbf{A}_{\tau}$ and offset $\mathbf{b}_{\tau}$ are determined by the tetrahedron's vertex positions and features and are constant within $\tau$.
Together with continuity across faces, this makes the encoding CPWA.
Composing it with affine layers and ReLU activations preserves this structure, enabling analytic mesh retrieval from the learned field, as described in Sec.~\ref{sec:tasks}.

The same vertex-based representation provides hash-free local support.
Each grid vertex stores its own learned feature vector, which can be updated independently.
For a fixed grid, changing one vertex feature produces
\begin{equation}
    \Delta\mathbf{e}(\mathbf{x})
    =\lambda_v\,\Delta\mathbf{f}_v,
    \label{eq:encoder_locality}
\end{equation}
where $\lambda_v$ is understood to be zero outside the tetrahedra containing $v$.
Changing the feature vector at vertex $v$ therefore affects the encoding only within tetrahedra incident to $v$.
With the decoder parameters fixed, modifying vertex features enables localized appearance editing without hash collisions causing changes in distant regions.

\paragraph{Grid adaptation and ray tracing.}
To represent coarse structures and fine details within a single encoding level, \ourmethod{} adapts local vertex capacity through densification, pruning, and vertex movement.

\emph{Densification} back-propagates gradients to the interpolated features to identify regions requiring additional capacity.
Let $\mathcal{L}_{\mathrm{fit}}$ denote the primary fitting objective: the RGB reconstruction loss $\mathcal{L}_{\mathrm{RGB}}$ for \textbf{T1} and \textbf{T3}, and the signed-distance loss $\mathcal{L}_{\mathrm{SDF}}$ for \textbf{T2}, defined in Eqs.~\eqref{eq:loss_rgb} and~\eqref{eq:loss_sdf}, respectively.
For each tetrahedron $\tau$, let $\mathcal{S}_{\tau}$ be a set containing the sampled points inside it whose interpolated features receive nonzero gradients.
The refinement score is
\begin{equation}
    \operatorname{score}(\tau)
    =
    \frac{\operatorname{diam}(\tau)}
         {|\mathcal{S}_{\tau}|}
    \sum_{\mathbf{x}\in\mathcal{S}_{\tau}}
    \left\|
        \frac{\partial\mathcal{L}_{\mathrm{fit}}}
             {\partial\mathbf{e}(\mathbf{x})}
    \right\|_1,
    \label{eq:encoder_refinement_score}
\end{equation}
where $\operatorname{diam}(\tau)$ denotes the length of the longest edge of $\tau$.
The score is zero when $\mathcal{S}_{\tau}$ is empty.
This criterion favors spatially coarse regions with strong feature gradients.
Insertion increases local vertex density and introduces independent feature vectors, initialized from the existing encoding as $\mathbf{f}_{\mathrm{new}} = \mathbf{e}_{\mathrm{old}}(\mathbf{p}_{\mathrm{new}})$, where $\mathbf{p}_{\mathrm{new}}$ and $\mathbf{f}_{\mathrm{new}}$ denote the position and feature vector of the inserted vertex, respectively, and $\mathbf{e}_{\mathrm{old}}$ is the encoding before insertion.

\emph{Pruning} identifies vertices whose features receive small training gradients over time and are similar to the mean feature vector of their neighbors.
These vertices become candidates for removal.
Removing selected vertices deletes their feature vectors and reduces local grid density.

\emph{Movement} redistributes existing capacity by optimizing vertex positions as $\mathbf{p}_v \leftarrow \mathbf{p}_v + \Delta\mathbf{p}_v$, where $\Delta\mathbf{p}_v$ is an Adam update computed from gradients propagated through the encoder.
Features remain attached to their vertices, so movement changes their spatial support without adding parameters.

Insertion and pruning trigger a full Delaunay rebuild over the active vertices.
During movement, connectivity is refreshed periodically or when geometric checks detect inverted or excessively compressed tetrahedra.
These refreshes update barycentric transforms and face adjacency while preserving surviving feature vectors.

The same connectivity supports tetrahedral ray tracing, our strategy for locating successive ray samples, illustrated in Fig.~\ref{fig:tetrahedral_traversal}.



\subsection{Tasks}
\label{sec:tasks}

\noindent \textbf{T1 -- \taskone}.
\ourmethod{} maps each normalized pixel coordinate $\mathbf{x}_i\in[0,1]^2$ to interpolated features $\mathbf{e}(\mathbf{x}_i)$.
An RGB decoder $D_{\mathrm{RGB}}$ predicts pixel colors as $\hat{\mathbf{c}}_i=D_{\mathrm{RGB}}(\mathbf{e}(\mathbf{x}_i))$.
We jointly optimize the encoder and decoder using the mean squared RGB loss:
\begin{equation}
    \mathcal{L}_{\mathrm{T1}}
    =\mathcal{L}_{\mathrm{RGB}}
    =\frac{1}{3|\mathcal{P}|}
    \sum_{i\in\mathcal{P}}
    \left\|\hat{\mathbf{c}}_i-\mathbf{c}_i\right\|_2^2,
    \label{eq:loss_rgb}
\end{equation}
where $\mathcal{P}$ indexes the sampled pixels and $\mathbf{c}_i$ is the reference RGB color of pixel $i$.\\

\noindent \textbf{T2 -- \tasktwo}.
\ourmethod{} supplies $\mathbf{e}(\mathbf{x})$ to a ReLU geometry decoder $D_{\mathrm{geom}}$, defining the CPWA scalar field $g(\mathbf{x})=D_{\mathrm{geom}}(\mathbf{e}(\mathbf{x}))$ and the predicted signed distance $\hat{s}(\mathbf{x})=\tanh(g(\mathbf{x}))$.
Given a batch $\mathcal{B}_{\mathrm{SDF}}$ of input points and reference signed distances $(\mathbf{x},s)$, we jointly optimize the encoder and decoder with a main objective combining L1 fitting and Eikonal regularization:
\begin{align}
    \mathcal{L}_{\mathrm{T2}}
    &=
    \mathcal{L}_{\mathrm{SDF}}
    +\lambda_{\mathrm{eik}}\mathcal{L}_{\mathrm{eik}},
    \label{eq:tasktwo_objective}\\
    \mathcal{L}_{\mathrm{SDF}}
    &=
    \frac{1}{|\mathcal{B}_{\mathrm{SDF}}|}
    \sum_{(\mathbf{x},s)\in\mathcal{B}_{\mathrm{SDF}}}
    |\hat{s}(\mathbf{x})-s|,
    \label{eq:loss_sdf}\\
    \mathcal{L}_{\mathrm{eik}}
    &=
    \frac{1}{|\mathcal{Q}|}
    \sum_{\mathbf{x}\in\mathcal{Q}}
    \left(\|\nabla_{\mathbf{x}}\hat{s}(\mathbf{x})\|_2-1\right)^2,
    \label{eq:loss_eik}
\end{align}
where $\mathcal{Q}$ is a set containing the points used for Eikonal regularization, and the weight $\lambda_{\mathrm{eik}}$ encourages unit SDF gradient norm.

Using the CPWA property described in Sec.~\ref{sec:adaptive_encoder}, \ourmethod{} supports analytic extraction directly over its single adaptive tetrahedral grid.
We subdivide each tetrahedron into convex cells $C$ with fixed ReLU activation. The affine form in Eq.~\eqref{eq:encoder_affine}  can thus be extended to the coupled encoder and decoder with coefficients $\mathbf{a}_C$ and $b_C$.
Since $\tanh$ is zero exactly when its input is zero, the surface $\hat{s}(\mathbf{x})=0$ is equivalently defined by $g(\mathbf{x})=0$, giving within each cell
\begin{equation}
    \{\mathbf{x}\in C\mid\hat{s}(\mathbf{x})=0\}
    =
    C\cap\{\mathbf{x}\mid\mathbf{a}_C^\top\mathbf{x}+b_C=0\}.
    \label{eq:extraction_surface_cell}
\end{equation}
We triangulate the resulting nondegenerate planar polygons and assemble them into the surface mesh.\\

\noindent \textbf{T3 -- \taskthree}.
\ourmethod{} forms the spatial representation of our image-supervised reconstruction pipeline, built on the volume-rendering framework of NeuS2~\cite{wang2023neus2}.
We feed $[\mathbf{x};\mathbf{e}(\mathbf{x})]$ to a ReLU geometry decoder following the formulation of \textbf{T2}, here using its CPWA scalar output directly as the SDF: $\hat{s}(\mathbf{x}) = g(\mathbf{x})$.
The geometry decoder produces the scalar SDF $g(\mathbf{x})$ together with auxiliary features for appearance prediction $k(\mathbf{x})$; we denote its full output vector by $\mathbf{h}(\mathbf{x})$.
A radiance decoder $D_{\mathrm{rad}}$ predicts local view-dependent radiance as
\begin{equation}
    \hat{\mathbf{L}}(\mathbf{x},\mathbf{d})
    =
    D_{\mathrm{rad}}\!\left(
        \mathbf{x},\mathbf{d},
        \mathbf{h}(\mathbf{x}),\boldsymbol{\nu}(\mathbf{x})
    \right),
    \label{eq:radiance_prediction}
\end{equation}
where $\mathbf{d}$ is the viewing direction, and $\boldsymbol{\nu}(\mathbf{x})$ denotes the normal-related features evaluated at position $\mathbf{x}$.
Let $j$ index sample points $\mathbf{x}_j$ along a camera ray, ordered from near to far, and write $\hat{s}_j=\hat{s}(\mathbf{x}_j)$.
The opacity of the interval between $\mathbf{x}_j$ and $\mathbf{x}_{j+1}$ is\begin{equation}
\begin{gathered}
    \alpha_j=
    \operatorname{clip}_{[0,1]}
    \left(
        1-\frac{\Phi_\beta(\hat{s}_{j+1})}
                {\Phi_\beta(\hat{s}_j)}
    \right),\\[4pt]
    \Phi_\beta(z)=\frac{1}{1+e^{-\beta z}},
\end{gathered}
\label{eq:sdf_opacity}
\end{equation}
where $\Phi_\beta$ is the logistic CDF and $\beta>0$ is a learned sharpness parameter.
Volume rendering accumulates the local radiance $\hat{\mathbf{L}}(\mathbf{x}_j,\mathbf{d})$ along each ray, weighted by opacity and transmittance, to obtain the rendered pixel color $\hat{\mathbf{c}}_i$.

We jointly optimize the encoder and decoders with the main objective
\begin{equation}
    \mathcal{L}_{\mathrm{T3}}
    =\mathcal{L}_{\mathrm{RGB}}
    +\lambda_{\mathrm{eik}}\mathcal{L}_{\mathrm{eik}},
    \label{eq:image_training_objective}
\end{equation}

The learned SDF remains CPWA, enabling analytic mesh extraction as in \textbf{T2}.\\

\noindent \textbf{T4 -- \taskfour}.
Starting from a trained scene as in \textbf{T3}, \ourmethod{} supports geometry and appearance editing on the tetrahedral encoder grid, without retraining.

\emph{Geometry editing.}
\ourmethod{} edits geometry by moving its grid vertices while keeping the learned features, the tetrahedral connectivity, and the geometry decoder fixed.
The learned field follows the grid deformation, producing a corresponding change in its zero-level surface and in the explicit mesh obtained through the analytic extraction described in \textbf{T2}.

\emph{Appearance editing.}
Since the geometry decoder produces both the SDF and $k(\mathbf{x})$, \ourmethod{} edits scene colors through its encoder grid.
We perturb the features at selected grid vertices to match the feature vector that represents the desired color.
The original features continue to determine the SDF and normals, preserving geometry while the edited features are used to render the edited color.

\section{Experiments and Results}
\label{sec:results}

This section presents the averaged per-task results and a dedicated ablation studies subsection. We use \hash{} for multiresolution hash encoding, \thash{} for tetrahedral multiresolution hash encoding and \mesh{} for methods that use a mesh-like scaffold for editing \cite{kulhanek2023tetra,bao2025neumeshplusplus}. Complete tables and additional ablations, quantitative and qualitative results for each task can be found in the supplementary material.

\suppressfloats[t]


\subsection{R1 -- \taskone}
\label{sec:toy_results}

\begin{table}[t]
\centering
\footnotesize
\setlength{\tabcolsep}{8pt}
\renewcommand{\arraystretch}{1.10}
\begin{tabular}{llcc}
\toprule
\textbf{Image} & \textbf{Method} &
\textbf{PSNR (dB)} $\uparrow$ &
\textbf{Parameters} $\downarrow$ \\
\midrule
\multirow{2}{*}{Albert}
& \hash & 50.31 & 5.75M \\
& \ourmethod & \cellcolor{firstbest}\textbf{59.81} & \cellcolor{firstbest}\textbf{3.61M} \\
\midrule
\multirow{2}{*}{Summer}
& \hash & 37.94 & 6.57M \\
& \ourmethod & \cellcolor{firstbest}\textbf{38.18} & \cellcolor{firstbest}\textbf{4.82M} \\
\midrule
\multirow{2}{*}{Pluto}
& \hash & 44.03 & 23.96M \\
& \ourmethod & \cellcolor{firstbest}\textbf{48.63} & \cellcolor{firstbest}\textbf{17.27M} \\
\midrule
\multirow{2}{*}{Girl}
& \hash & 35.96 & 86.20M \\
& \ourmethod & \cellcolor{firstbest}\textbf{36.14} & \cellcolor{firstbest}\textbf{58.74M} \\
\bottomrule
\end{tabular}
\caption{\textbf{R1} -- Image fitting comparison with Hash Encoding (Instant-NGP).
Our method consistently achieves higher PSNR with fewer parameters.}
\label{tab:image_fitting}
\end{table}

We evaluate image fitting on four high resolution RGB images at their original resolutions, up to $8000\times8000\times3$ pixels. We optimize the weights of a single hidden unit (size $64$) MLP to fit each image using Adam for $350$ epochs and learning rate $0.01$. Each 2D sample is mapped to a $32$-dimensional input vector using \ourmethod. 

\noindent  \ourmethod{} outperforms \hash{} \cite{muller2022instant} in terms of PSNR, while simultaneously reducing the number of required parameters (Table \ref{tab:image_fitting}). Qualitative results versus \hash{} are shown in Fig. \ref{fig:qualitative_2D}.


\subsection{\textbf{R2} -- \tasktwo}
\label{sec:pc2mesh_results}
We evaluate analytic surface reconstruction on five meshes from the Stanford 3D Scanning Repository. For each normalized mesh, we generate 500,000 SDF supervision samples, with $95\%$ perturbed surface samples and $5\%$ uniformly sampled points in the unit cube $[-1,1]^3$. Surface samples are perturbed with isotropic Gaussian noise of standard deviation $\sigma=0.0333$. Each query point is mapped by \ourmethod{} to an 8-dimensional feature vector and decoded by an MLP with three hidden layers of $12$ neurons each. We train for $10$ epochs using batches of $1,000$ samples, and Adam with an initial decoder learning rate of $0.001$ and cosine annealing. We report our results against TetraSDF Large \cite{tetrasdf} in Table \ref{tab:stanford_comparison}, showing how \ourmethod{} can outperform tetrahedral hash encoding in terms of Chamfer Distance (CD), extraction time, number of parameters, SSDF and VSDF (as defined in \cite{tetrasdf}) while improving the output topology, with a greater number of Delaunay edges and less skinny triangles (Fig. \ref{fig:qualitative_3D}).

\begin{table}[t]
\centering
\small
\setlength{\tabcolsep}{5pt}
\renewcommand{\arraystretch}{1.10}

\resizebox{.4\textwidth}{!}{%
\begin{tabular}{llcccccc}
\toprule
\textbf{Metric}
& \textbf{Method}
& \textbf{Mean} \\
\midrule

\multirow{2}{*}{CD $\downarrow$}
& TetraSDF L (\thash)
& 1.718 \\

& Ours (\ourmethod)
& \cellcolor{firstbest}\textbf{1.611} \\

\midrule

\multirow{2}{*}{Extract (s) $\downarrow$}
& TetraSDF L (\thash)
& 2.859 \\

& Ours (\ourmethod)
& \cellcolor{firstbest}\textbf{0.446} \\

\midrule

\multirow{2}{*}{Params. (M) $\downarrow$}
& TetraSDF L (\thash)
& 16.778 \\

& Ours (\ourmethod)
& \cellcolor{firstbest}\textbf{5.221} \\

\midrule

\multirow{2}{*}{Delaunay edges (\%) $\uparrow$}
& TetraSDF L (\thash)
& 82.37 \\

& Ours (\ourmethod)
& \cellcolor{firstbest}\textbf{83.50} \\

\midrule

\multirow{2}{*}{Skinny triangles (\%) $\downarrow$}
& TetraSDF L (\thash)
& 19.55 \\

& Ours (\ourmethod)
& \cellcolor{firstbest}\textbf{9.14} \\

\midrule

\multirow{2}{*}{SSDF $\downarrow$}
& TetraSDF L (\thash)
& 0.076 \\

& Ours (\ourmethod)
& \cellcolor{firstbest}\textbf{0.027} \\

\midrule

\multirow{2}{*}{VSDF $\downarrow$}
& TetraSDF L (\thash)
& 0.078 \\

& Ours (\ourmethod)
& \cellcolor{firstbest}\textbf{0.027} \\

\bottomrule
\end{tabular}%
}

\caption{\textbf{R2} -- Surface reconstruction on the Stanford dataset.
We compare reproduced TetraSDF (large) with our method using
identical geometry supervision and 5,000 training updates.
We report Chamfer Distance (CD, $\times 10^{-3}$),
analytic extraction time in seconds, allocated parameters in millions,
intrinsic Delaunay edge compliance (\%), and skinny triangles (\%).}
\label{tab:stanford_comparison}
\end{table}



\begin{table*}[t]
\centering
\small
\setlength{\tabcolsep}{5pt}
\renewcommand{\arraystretch}{1.05}

\begin{tabular}{lccccccccc}
\toprule
& \multicolumn{3}{c}{\textbf{NeRF Synthetic}}
& \multicolumn{3}{c}{\textbf{DTU}}
& \multicolumn{3}{c}{\textbf{Mean}} \\
\cmidrule(lr){2-4}
\cmidrule(lr){5-7}
\cmidrule(lr){8-10}
\textbf{Method}
& PSNR $\uparrow$
& SSIM $\uparrow$
& CD $\downarrow$
& PSNR $\uparrow$
& SSIM $\uparrow$
& CD $\downarrow$
& PSNR $\uparrow$
& SSIM $\uparrow$
& CD $\downarrow$ \\
\midrule

Tetra-NeRF \cite{kulhanek2023tetra} (\mesh)
& 29.71
& 0.948
& 14.24
& 27.09
& 0.905
& 2.65
& 28.04
& 0.927
& 8.35 \\

NeuMesh++ \cite{bao2025neumeshplusplus} (\mesh)
& 32.32
& 0.955
& --
& 28.39
& 0.940
& --
& 30.36
& 0.9475
& -- \\

NeuS2 \cite{wang2023neus2} (\hash)
& 31.58
& 0.953
& \cellcolor{secondbest}4.10
& 32.14
& 0.938
& \cellcolor{secondbest}0.711
& 31.86
& 0.9455
& \cellcolor{secondbest}2.4055 \\

Instant-NGP \cite{muller2022instant} (\hash)
& \cellcolor{secondbest}32.75
& \cellcolor{secondbest}0.960
& 4.71
& \cellcolor{secondbest}32.27
& \cellcolor{secondbest}0.942
& 1.77
& \cellcolor{secondbest}32.51
& \cellcolor{secondbest}0.9510
& 3.2400 \\

Ours (\ourmethod)
& \cellcolor{firstbest}33.27
& \cellcolor{firstbest}0.969
& \cellcolor{firstbest}3.93
& \cellcolor{firstbest}32.34
& \cellcolor{firstbest}0.943
& \cellcolor{firstbest}0.705
& \cellcolor{firstbest}\textbf{32.81}
& \cellcolor{firstbest}\textbf{0.9560}
& \cellcolor{firstbest}\textbf{2.3725} \\

\bottomrule
\end{tabular}
\caption{\textbf{R3} -- Novel-view synthesis appearance and reconstruction accuracy
on NeRF Synthetic and DTU.
Mean reports the arithmetic average across the two datasets.
CD is reported in millimeters.
}
\label{tab:nvs_comparison}
\end{table*}

\begin{figure}[t]
    \centering
    \includegraphics[width=\linewidth]{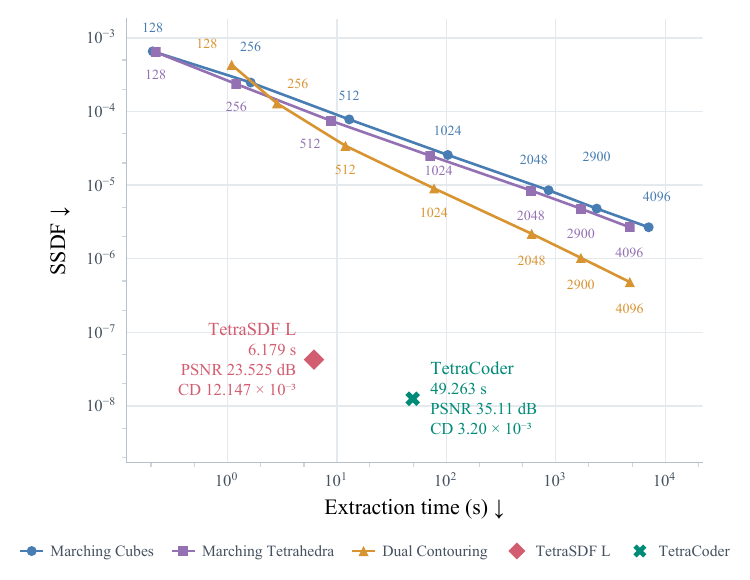}
    \caption{\textbf{R3} -- Analytic extraction and grid-based extractors on the same frozen Lego field.
    SSDF measures mesh--field agreement.}
    \label{fig:ssdf_extraction_time}
\end{figure}

\subsection{\textbf{R3} -- \taskthree}
\label{sec:3dnvs_results}

We evaluate both novel view synthesis from images and 3D analytic mesh reconstruction from the learned volume using PSNR, SSIM and CD. We encode 3D points to $28$-dimensional features using \ourmethod{} and decode them using a decoder matching the structure of the NeuS2 \cite{wang2023neus2} SDF and radiance decoders, training for $20k$ iterations on NeRF Synthetic \cite{mildenhall2020nerf} and DTU \cite{6909453}. The choice of the $28$ features is not arbitrary: NeuS2 uses $14$ levels of multi-resolution hash encoding with $2$ features per level. Thus, we choose $28$ features to match the same architectural capacity.

Table \ref{tab:nvs_comparison} shows \ourmethod{} consistently outperforming similar methods using \hash{} \cite{wang2023neus2, muller2022instant}, as well as recent methods specifically designed with additional editing scaffolds \cite{kulhanek2023tetra, bao2025neumeshplusplus}. Qualitative results are shown in Fig. \ref{fig:qualitative_nvs}.

Bridging the gap with task \textbf{T2} (Sec.~\ref{sec:pc2mesh_results}), we adapt TetraSDF \cite{tetrasdf} to fit the same image-based task. We first train this new architecture with a decoder matching the shape and size of TetraSDF L: a $4$-levels encoder with a feature size of $2$, combined with a $(8, 128)$ MLP. We then train the same modified architecture but with a significantly larger, $14$-levels NeuS2-style encoder and decoder, which we name TetraSDF-NeuS for convenience.

Fig. \ref{fig:ssdf_extraction_time} shows a logarithmic plot comparing the analytic extraction time versus the SSDF for different methods. Classical methods (Marching Cubes, Marching Tetrahedra, Dual Contouring), show an exponential increase in extraction time when scaling the grid size. In the same plot, the modified TetraSDF L achieves similarly low SSDF to TetraCoder ($4.26\times10^{-8}$ versus $1.25\times10^{-8}$), consistent with both methods using analytic extraction. However, it reaches only $23.53$\,dB PSNR and produces $93{,}065$ boundary edges, $107{,}547$ non-manifold edges and $26{,}374$ duplicate triangles (all $0$ for \ourmethod{}), showing that its limited capacity is insufficient to learn a high-quality SDF from images, yielding a severely defective mesh. At the same time, scaling TetraSDF-NeuS to NeuS2-like capacity makes construction of the polyhedral complex prohibitively expensive in memory and runtime, limiting its practical use at that scale. As shown in Fig. \ref{fig:ssdf_extraction_time}, in less than 50 seconds \ourmethod{} achieves the lowest SSDF among all the candidates, almost two orders of magnitude lower than Dual Contouring trained on a $4096 \times 4096 \times 4096$ grid for 2.5 hours. At resolution $2900 \times 2900 \times 2900$ the other grid-based methods roughly match the resolution and storage size of our analytic mesh.


\begin{table}[t]
\centering
\setlength{\tabcolsep}{5pt}
\renewcommand{\arraystretch}{1.10}

\resizebox{.45\textwidth}{!}{%
\begin{tabular}{ccccc}
\toprule
\multirow{2}{*}{\textbf{Editing \%}}
& \multicolumn{2}{c}{\textbf{2D MSE$\downarrow$} ($\times 10^{-2}$)}
& \multicolumn{2}{c}{\textbf{3D MSE$\downarrow$}} \\
\cmidrule(lr){2-3}
\cmidrule(lr){4-5}
& \ourmethod & \hash & \ourmethod & \hash \\
\midrule

20
& \cellcolor{firstbest}0.048
& 0.536
& \cellcolor{firstbest}0.000
& 0.055 \\

40
& \cellcolor{firstbest}0.040
& 0.655
& \cellcolor{firstbest}0.001
& 0.139 \\

60
& \cellcolor{firstbest}0.043
& 0.942
& \cellcolor{firstbest}0.001
& 0.091 \\

80
& \cellcolor{firstbest}0.160
& 2.221
& \cellcolor{firstbest}0.003
& 0.155 \\

\bottomrule
\end{tabular}%
}

\caption{\textbf{R4} -- Before--after editing MSE on the unpainted region for 2D and 3D scenes at different amounts of editing.}
\label{tab:editing_mse}
\end{table}

\subsection{\textbf{R4} -- \taskfour}
\label{sec:editing_results}

We evaluate 3D scene editing both in terms of geometry and appearance, and report the accuracy and the speed of our editing process. Table~\ref{tab:editing_mse} quantifies unintended appearance changes by reporting the before--after MSE in unpainted regions.
\ourmethod{} consistently preserves these regions better than \hash{}. These results demonstrate that local edits introduce limited changes outside the intended region.

\begin{figure*}[t]
    \centering

    \newcommand{\twodrow}{%
        \includegraphics[width=\linewidth]{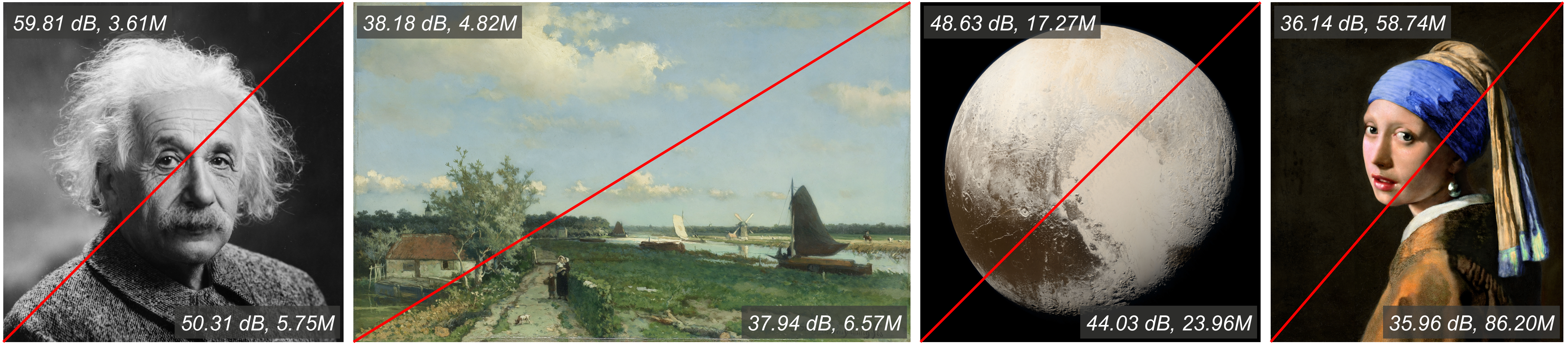}%
    }
    
    \newcommand{\threedrow}{%
        \includegraphics[width=.2\linewidth]{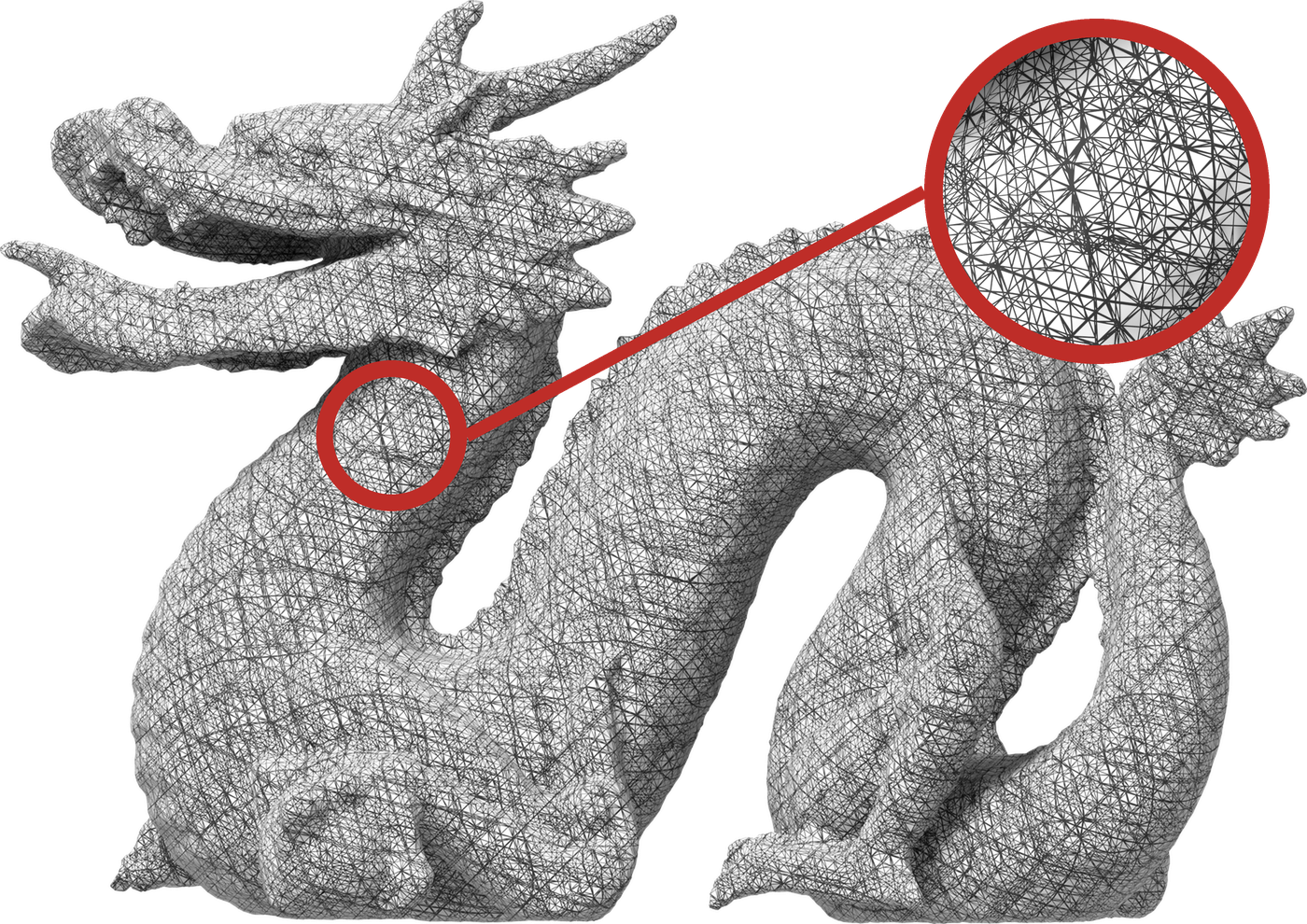}\hfill
        \includegraphics[width=.2\linewidth]{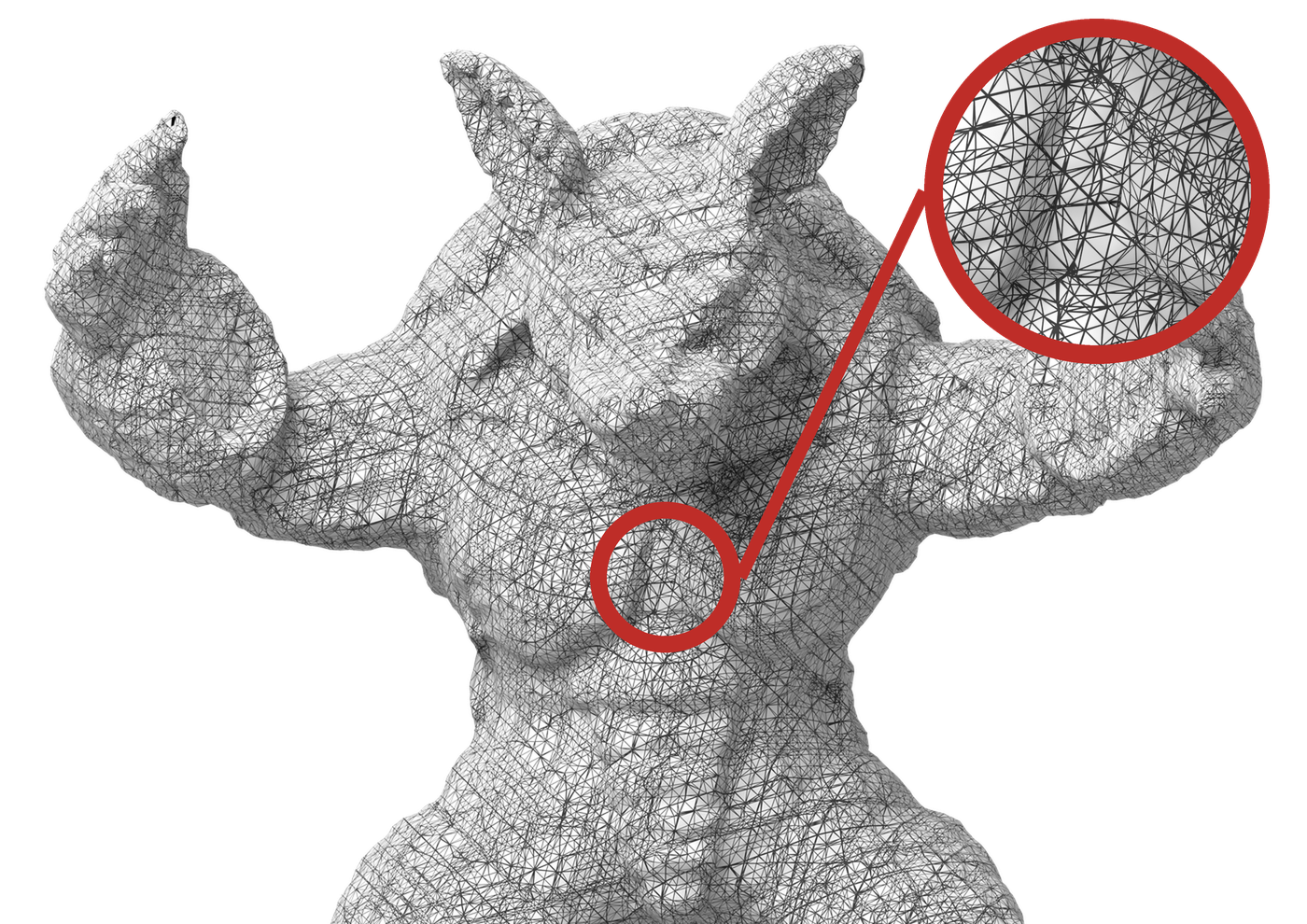}\hfill
        \includegraphics[width=.2\linewidth]{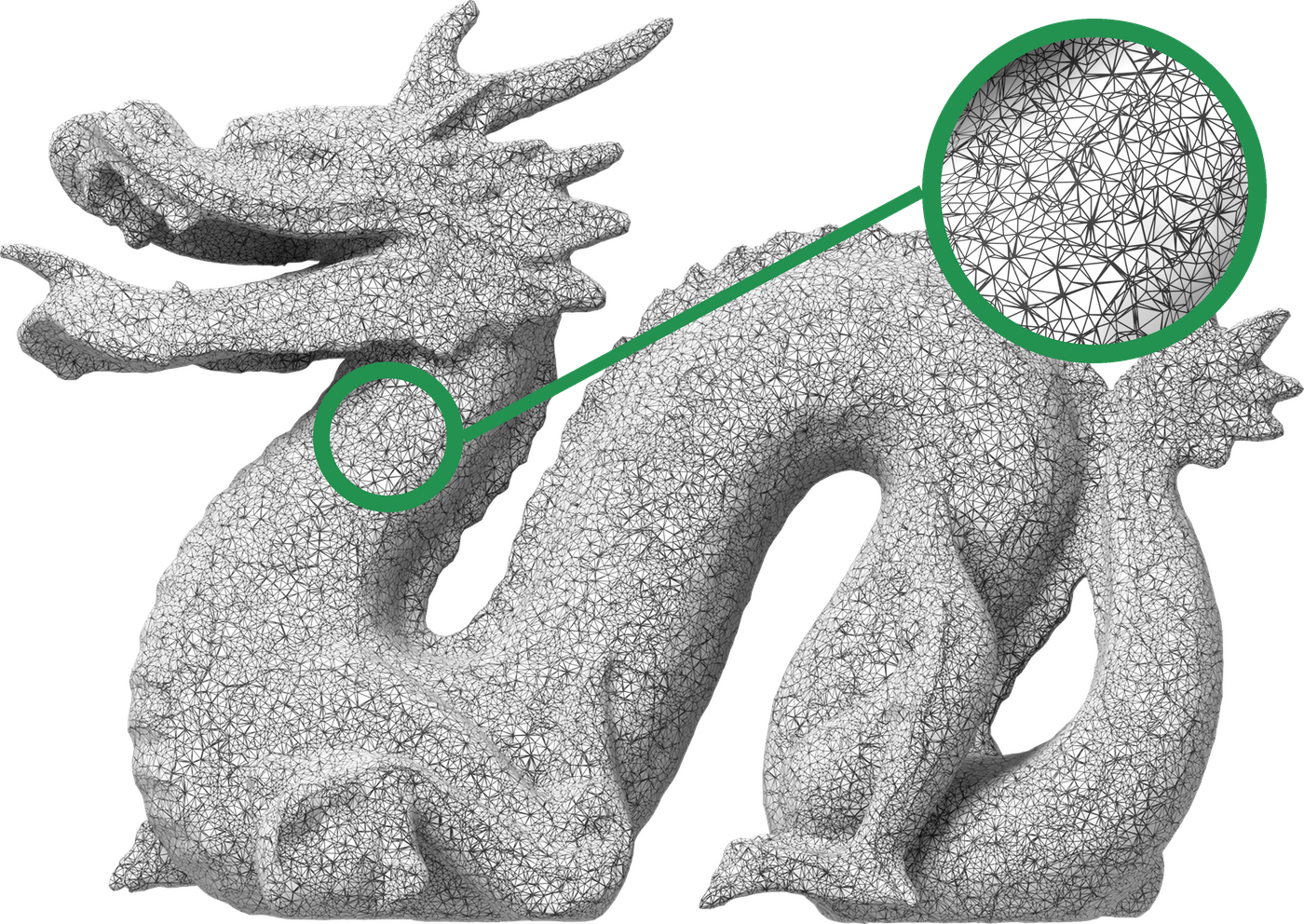}\hfill
        \includegraphics[width=.2\linewidth]{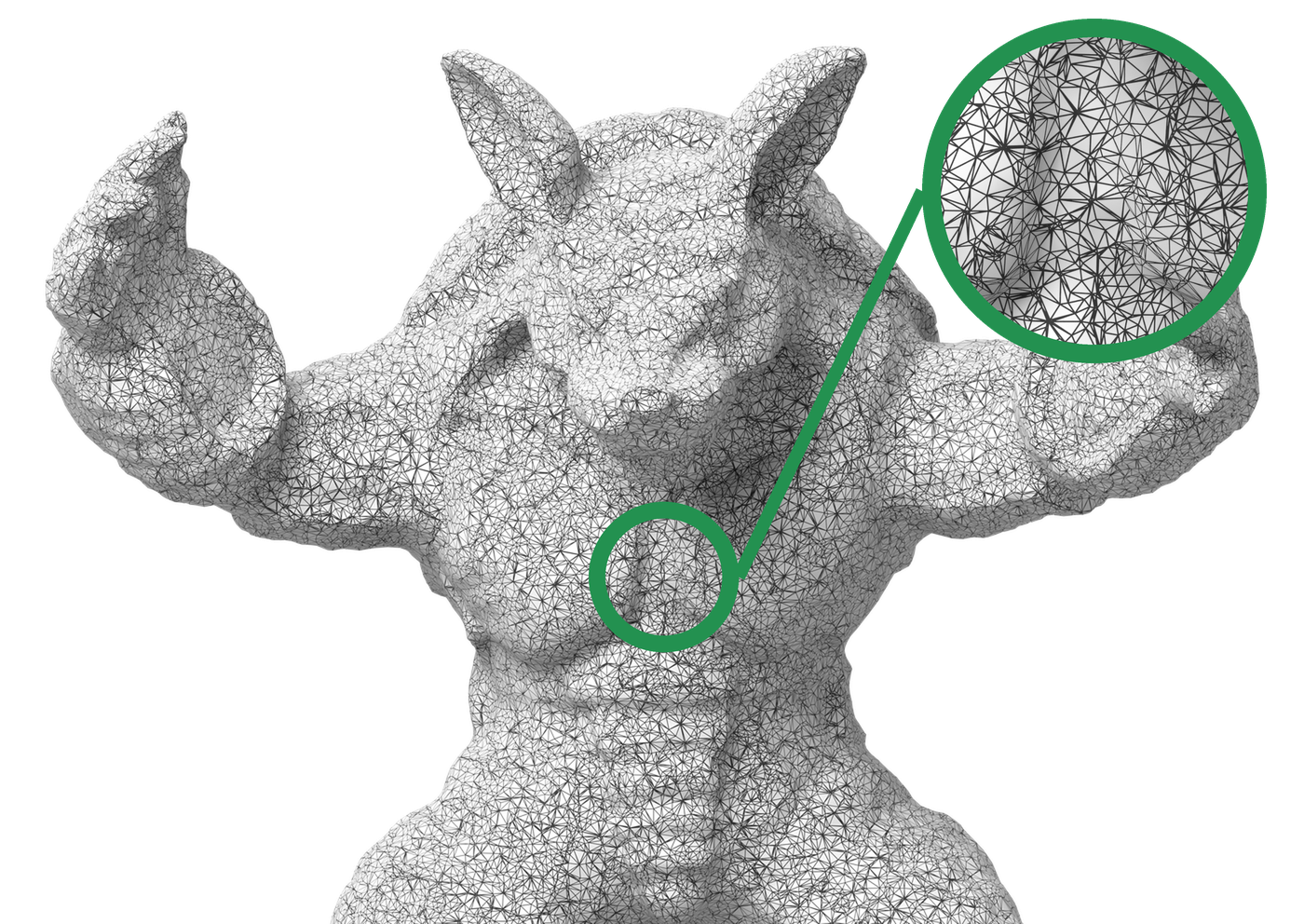}%
    }
    
    \newcommand{\nvsrow}{%
        \includegraphics[width=.2\linewidth]{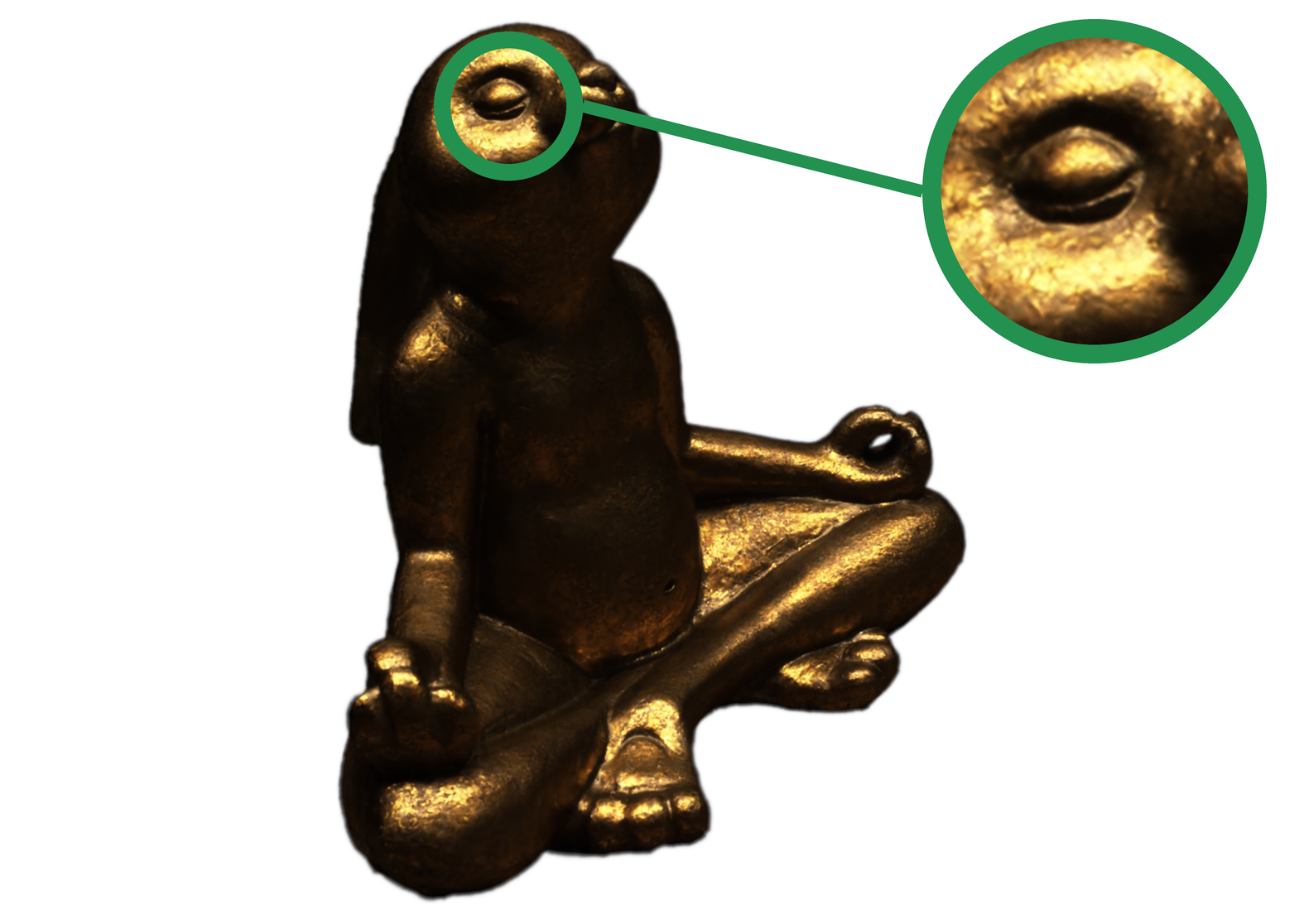}\hfill
        \includegraphics[width=.2\linewidth]{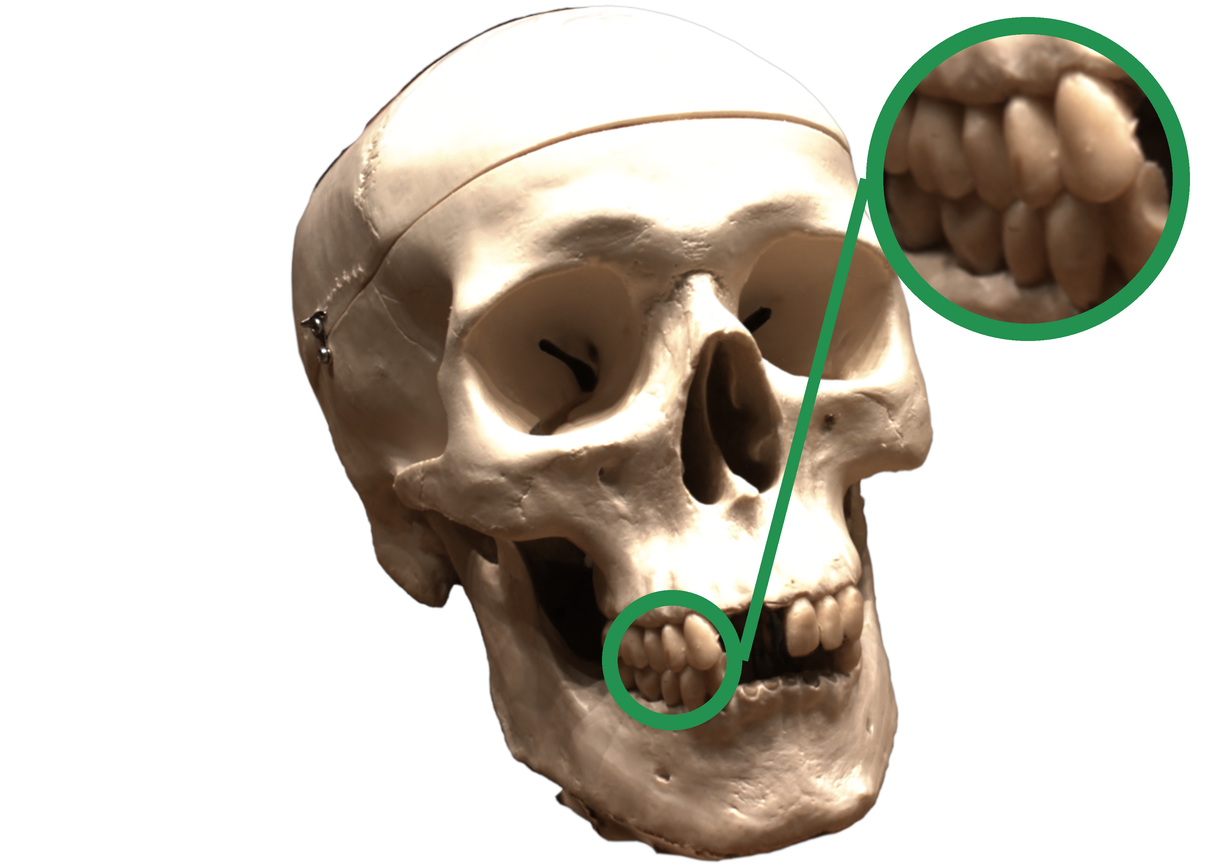}\hfill
        \includegraphics[width=.2\linewidth]{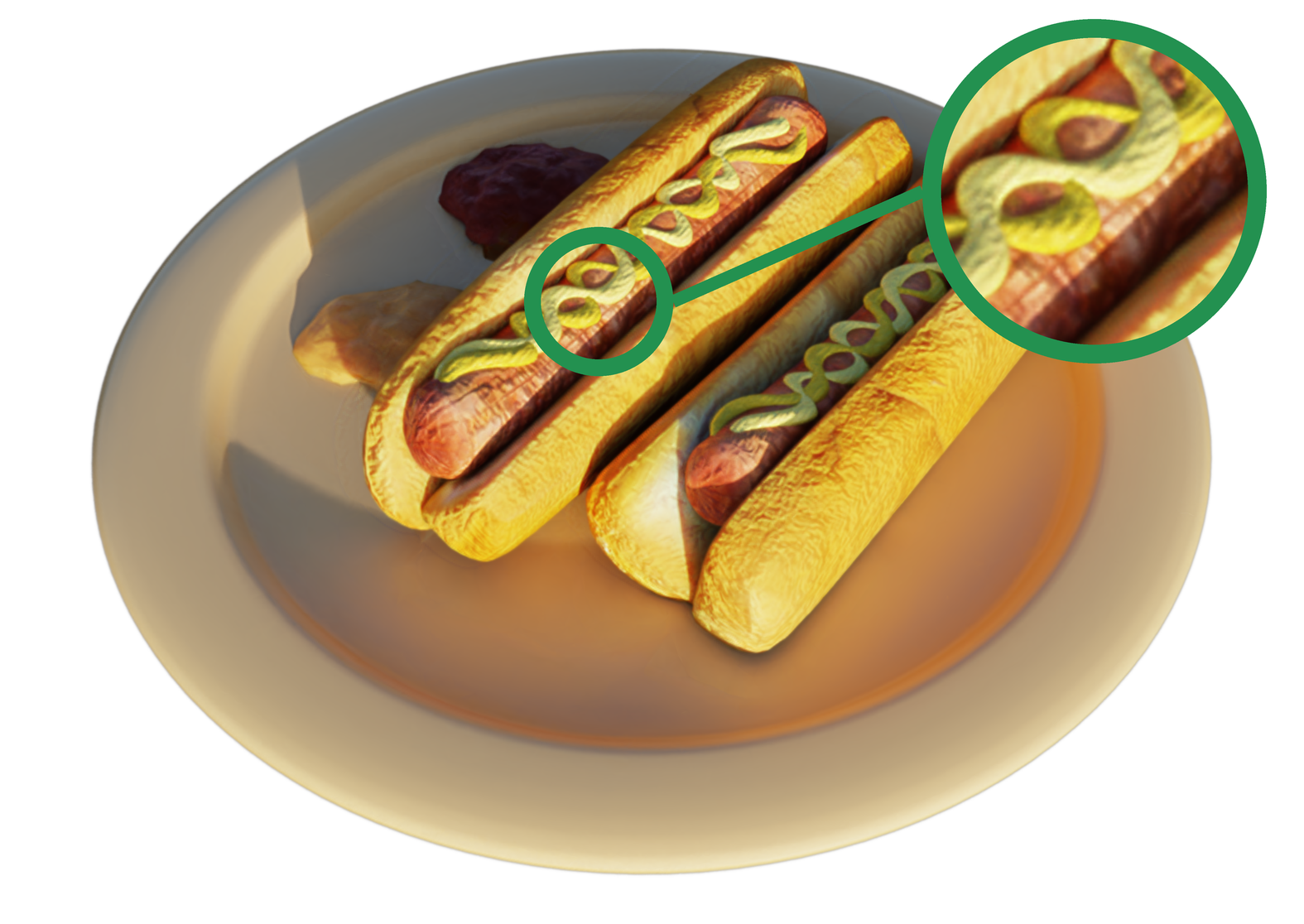}\hfill
        \includegraphics[width=.2\linewidth]{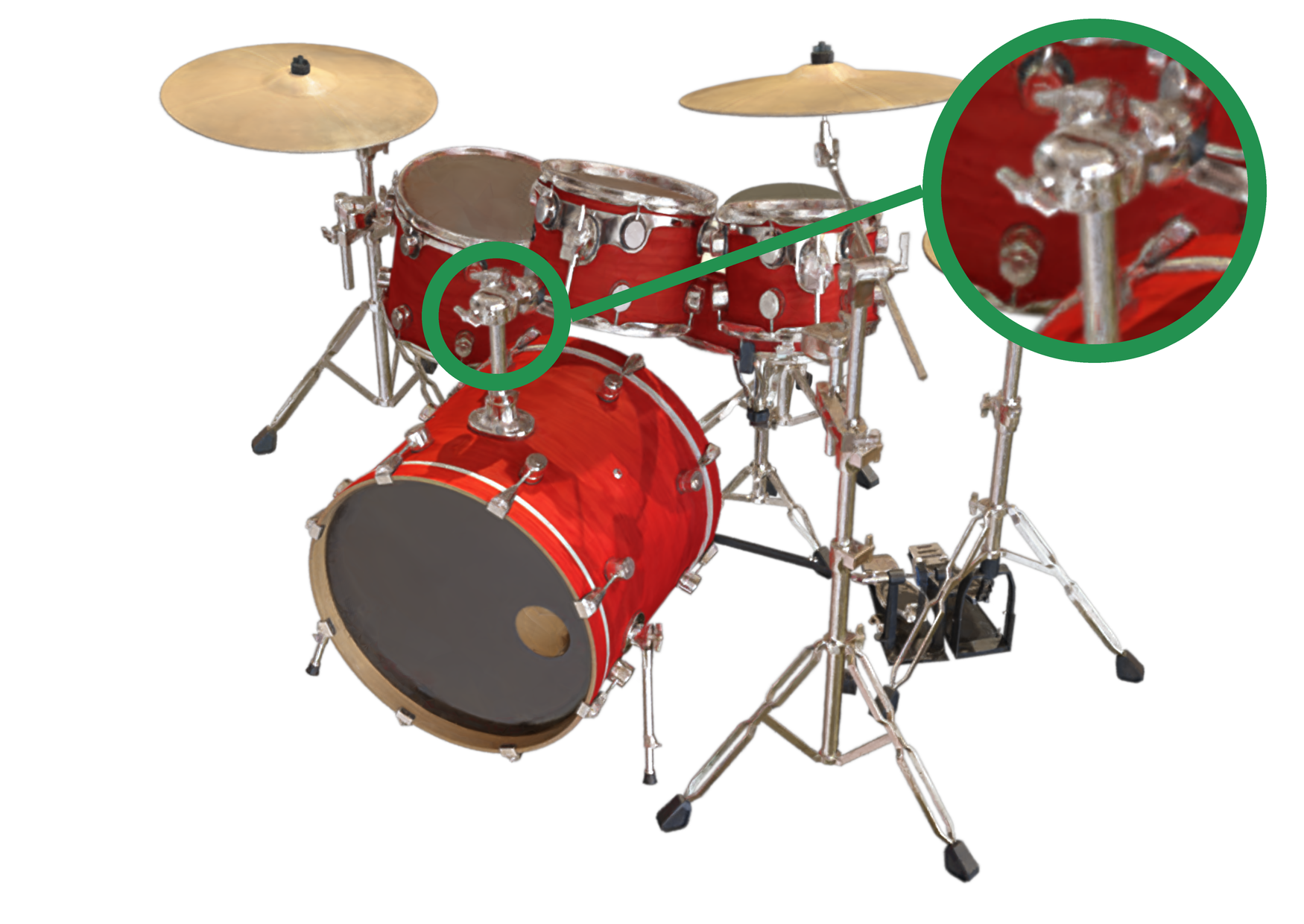}%
    }
    
    \newcommand{\editingrow}{%
        \includegraphics[width=.2\linewidth]{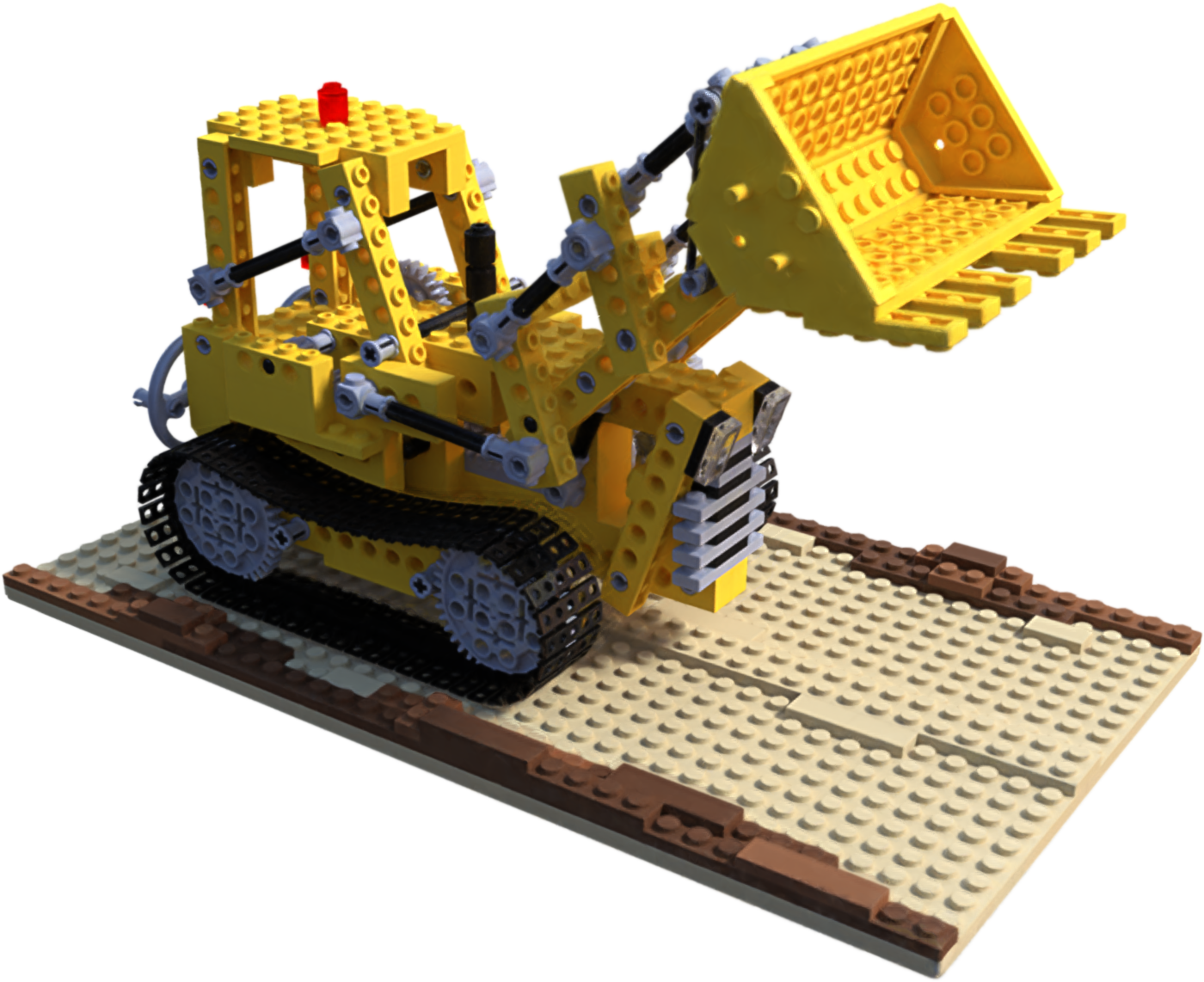}\hfill
        \includegraphics[width=.2\linewidth]{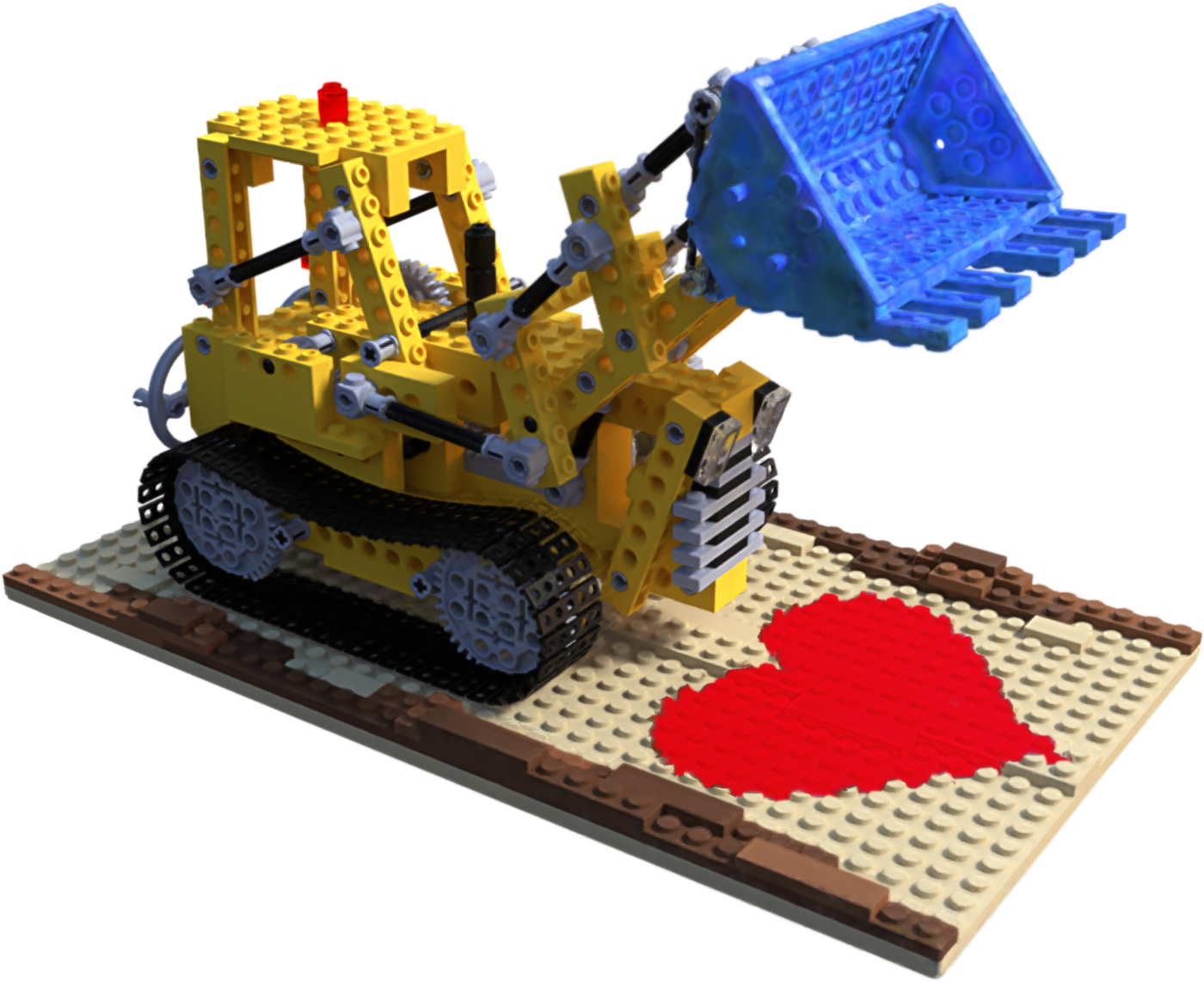}\hfill
        \includegraphics[width=.2\linewidth]{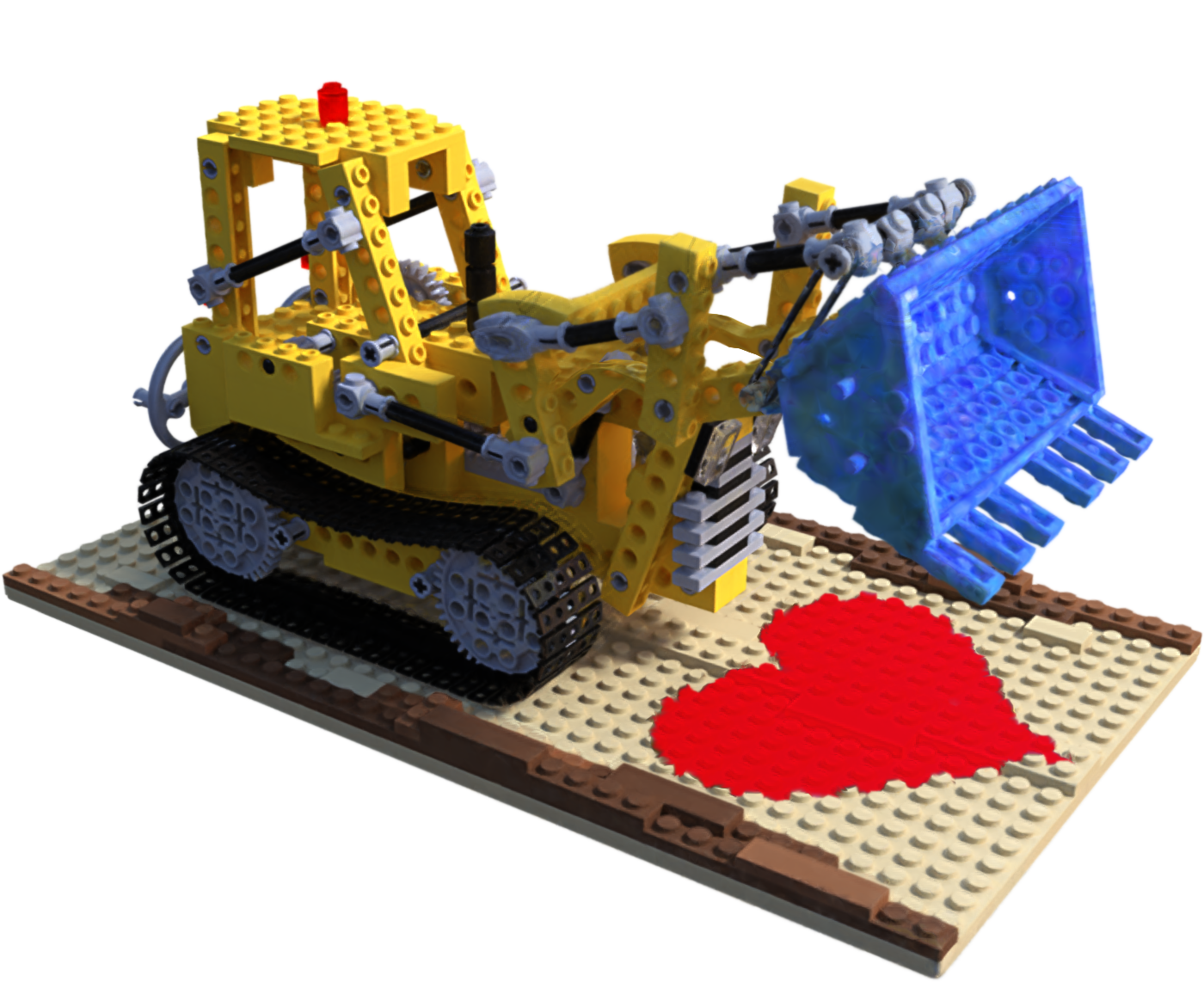}\hfill
        \includegraphics[width=.2\linewidth]{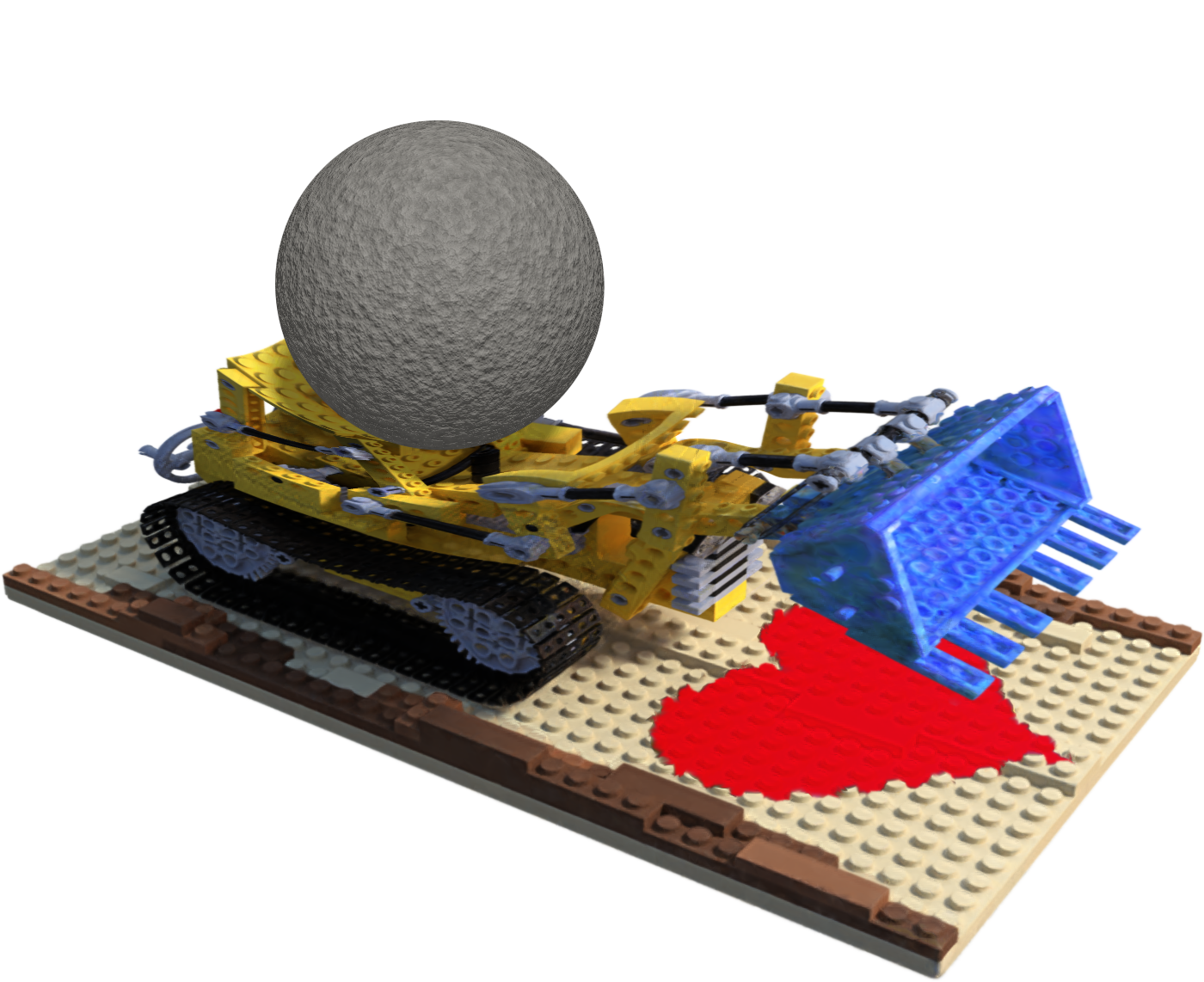}%
    }

    \begin{subfigure}{\textwidth}
        \centering
        \twodrow
        \caption{\textbf{R1} -- \textbf{2D Image fitting} -- Top-left: \ourmethod{}. Bottom-right: \hash{}. Compared to \hash{}, we obtaing greater PSNR values with fewer parameters.}
        \label{fig:qualitative_2D}
    \end{subfigure}
    \par\medskip

    \begin{subfigure}{\textwidth}
        \centering
        \threedrow
        \caption{\textbf{R2} -- \textbf{3D Point cloud to Mesh} -- Details in \textbf{\textcolor{red}{red}}: TetraSDF artifacts due to the multi-level polyhedral extraction. Details in \textbf{\textcolor{teal}{green}}: \ourmethod{}.}
        \label{fig:qualitative_3D}
    \end{subfigure}
    \par\medskip

    \begin{subfigure}{\textwidth}
        \centering
        \nvsrow
        \caption{\textbf{R3} -- \textbf{3D Novel view synthesis} -- Details in \textbf{\textcolor{teal}{green}}: \ourmethod{} can model view-dependent effects. Left: DTU Dataset. Right: Nerf Synthetic Dataset.}
        \label{fig:qualitative_nvs}
    \end{subfigure}
    \par\medskip

    \begin{subfigure}{\textwidth}
        \centering
        \editingrow
        \caption{\textbf{R4} -- \textbf{Editing} -- Left to right:
        \ourmethod{} fitted scene, appearance editing,
        geometric editing, and physics simulation involving
        another 3D object.}
        \label{fig:qualitative_editing}
    \end{subfigure}

    \caption{Qualitative results on \textbf{T1}: \taskone, \textbf{T2}: \tasktwo, \textbf{T3}: \taskthree, \textbf{T4}: \taskfour. \ourmethod{} can adapt to different decoders, tackling multiple tasks with a single unified representation.}
    \label{fig:qualitative}
\end{figure*}

Fig.~\ref{fig:qualitative_editing} illustrates appearance editing, geometric deformation, and interaction with another object through physics simulation. Compared with NeuMesh++~\cite{bao2025neumeshplusplus}, \ourmethod{} achieves approximately $2\times$ the rendering speed and $19\times$ and $18\times$ lower geometry and appearance editing latencies, respectively. Additional details can be found in the supplementary results.




\begin{table}[t]
\centering
\small
\setlength{\tabcolsep}{5pt}
\renewcommand{\arraystretch}{1.10}
\newcommand{\ablationlastwidth}{\dimexpr\columnwidth-12\tabcolsep-0.07\columnwidth-0.07\columnwidth-0.07\columnwidth-0.14\columnwidth-0.19\columnwidth-0.01pt\relax}
\begin{tabular}{
*{3}{>{\centering\arraybackslash}m{0.07\columnwidth}}
>{\centering\arraybackslash}m{0.14\columnwidth}
>{\centering\arraybackslash}m{0.19\columnwidth}
>{\centering\arraybackslash}m{\ablationlastwidth}}
\toprule
\textbf{D} & \textbf{P} & \textbf{M}
& \textbf{CD} $\downarrow$
& \textbf{PSNR} $\uparrow$
& \textbf{SSIM} $\uparrow$ \\
\midrule
\encodingcross & \encodingcross & \encodingcheck & 7.185 & 31.5678 & 0.959210 \\
\encodingcross & \encodingcheck & \encodingcross & 7.929 & 31.2010 & 0.955343 \\
\encodingcross & \encodingcheck & \encodingcheck & 9.036 & 31.5586 & 0.959147 \\
\encodingcheck & \encodingcross & \encodingcheck & \cellcolor{secondbest}6.534 & 33.1690 & 0.968326 \\
\encodingcheck & \encodingcheck & \encodingcross & 7.094 & \cellcolor{secondbest}33.2307 & \cellcolor{secondbest}0.968409 \\
\encodingcheck & \encodingcheck & \encodingcheck & \cellcolor{firstbest}\textbf{3.926} & \cellcolor{firstbest}\textbf{33.2647} & \cellcolor{firstbest}\textbf{0.968904} \\
\bottomrule
\end{tabular}

\caption{Site-update ablation on NeRF Synthetic.
D, P, and M denote densification, pruning, and site movement, respectively;
ticks and crosses indicate enabled and disabled operations.
We report mean CD ($\times 10^{-3}$, $\downarrow$),
PSNR (dB, $\uparrow$), and SSIM ($\uparrow$).}
\label{tab:site_ablation}
\end{table}

\subsection{Ablation studies}
\label{sec:ablation_studies}

We ablate densification (\textbf{D}), pruning (\textbf{P}), and site movement (\textbf{M}) on NeRF Synthetic (Table \ref{tab:site_ablation}). The complete configuration performs best across all metrics. Densification mainly supports appearance quality, while pruning and site movement primarily improve geometric accuracy. Their removal causes substantial geometric degradation with limited photometric changes, suggesting that similar rendered images can hide different surfaces and emphasizing the importance of site organization beyond appearance fitting.

\section{Conclusion}
\label{sec:conclusion}


We presented \ourmethod{}, a single-resolution, hash-free adaptive tetrahedral encoding that bridges implicit and explicit representations of geometry. By combining learnable vertex features with an adaptive spatial discretization, our approach efficiently allocates representational capacity while avoiding the limitations of multiresolution grids and hash collisions. Its continuous piecewise-affine structure enables exact analytic mesh extraction, establishing a direct correspondence between the learned implicit surface and its explicit representation. 
The presented results establish \ourmethod{} as a versatile spatial encoding that unifies neural field learning, explicit surface reconstruction, and scene manipulation within a single adaptive representation. To address the current limitations, future work will extend \ourmethod{} to dynamic 4D scenes, enable analytic mesh extraction at multiple levels of detail, and prove the bidirectionality of the implicit-explicit link by recovering the learned SDF from the extracted mesh.

\section{Acknowledgements}
\label{sec:acknowledgements}
This research was supported by grants from NVIDIA and utilized two NVIDIA RTX PRO 6000 Blackwell Max-Q Workstation Edition GPUs.

{
    \small
    \bibliographystyle{ieeenat_fullname}
    \bibliography{main}
}

\end{document}